\documentclass[11pt]{article}
\usepackage[a4paper, margin = 1in]{geometry}
\usepackage[T1]{fontenc}
\usepackage{lmodern}
\usepackage[sorting = anyt, maxbibnames=9,maxcitenames=2, backend = bibtex, style=authoryear, doi=false, url=true, dashed=false]{biblatex}
\usepackage[titletoc,page]{appendix}

\renewbibmacro{in:}{}
\usepackage{graphicx}
\usepackage{setspace}
\usepackage{gensymb}
\usepackage{enumerate}
\usepackage{kotex}
\usepackage{abstract,lipsum}
\usepackage[colorlinks=true,linkcolor=blue,citecolor=blue,urlcolor=blue]{hyperref}

\usepackage{titlesec}
\titleformat{\section}[block]{\normalfont\Large\bfseries}{\thesection}{1.0em}{}
\titleformat{\subsection}[block]{\normalfont\Large\bfseries}{\thesubsection}{1.0em}{}

\usepackage{amsmath}
\usepackage{amssymb}
\usepackage{amsthm}
\usepackage{amsfonts}
\usepackage{upgreek}
\usepackage{mathtools}
\usepackage{mathrsfs}
\usepackage{bm}
\newcommand\defeq{\stackrel{\mathclap{\normalfont\tiny\mbox{def}}}{=}}
\usepackage[symbol]{footmisc}

\usepackage{color}
\usepackage{multirow}
\usepackage{algorithm,algorithmic}
\usepackage{verbatim}

\usepackage{tcolorbox}
\usepackage[english]{babel}
\theoremstyle{remark}

\begin{document}

\begin{center}
       \fontsize{15pt}{15pt}\selectfont \textbf{Mechanical anisotropy of 3D-printed digital materials at large strains}
       
       \vspace*{0.3in}
       \fontsize{9.5pt}{9.5pt}\selectfont Seunghwan Lee$^{1}$, Gisoo Lee$^{2}$, Seounghee Yun$^{1}$, Sumin Lee$^{2}$, Jeonyoon Lee$^{2}$, Hansohl Cho$^{1\dagger}$  \\
       \vspace*{0.3in}
       \fontsize{9pt}{9pt}\selectfont $^{1}$Department of Mechanical Engineering, $^{2}$Department of Aerospace Engineering, Korea Advanced Institute of Science and Technology, Daejeon, 34141, Republic of Korea \\ 

\vspace*{0.2in}
\fontsize{9.5pt}{9.5pt}\selectfont E-mail: $^\dagger$hansohl@kaist.ac.kr
\end{center}

\renewenvironment{abstract}
{\small 
\noindent \rule{\linewidth}{.5pt}\par{\noindent \bfseries \abstractname.}}
{\medskip\noindent \rule{\linewidth}{.5pt}
}

\vspace*{0.3in}
\onehalfspacing
\begin{abstract}
\fontsize{10pt}{10pt}\selectfont
3D-printed digital materials whose mechanical behavior travels between those from thermoplastic to rubbery polymers have become increasingly important. However, their mechanical functionalities have not been fully exploited due to intrinsic mechanical anisotropy resulting from microstructural heterogeneity. Here, we combine mechanical testing, microscopy analysis and micromechanical modeling for a comprehensive understanding of complex deformation mechanisms responsible for the printing-orientation-dependent nonlinear mechanical behavior of digital materials at small to large strains. Towards this end, we construct representative volume elements that account for highly anisotropic microstructural features resulting from the printing-orientation-dependent diffusion and mixing between photocurable base resins. We then demonstrate, through micromechanical analysis, that stable compressive deformation of well-aligned elliptical hard thermoplastic inclusions embedded within the surrounding soft rubbery matrix gives rise to initial elastic anisotropy. Our experimental and micromechanical modeling results also show that the interplay between buckling instability and plastic deformation of the high-aspect-ratio hard domains governs mechanical anisotropy at large strains as well as the printing-orientation-dependent resilience and energy dissipation capabilities in these digital materials.
\\
\end{abstract} 

\doublespacing
\section{Introduction}
Understanding processing-microstructure-property relationships is of critical importance in three-dimensional (3D) printing (\cite{gibson2021additive, gu2021material, park20223d}). It has become increasingly important with the growing interest in multi-material printing (\cite{zhu2024toward, brown2025multimaterial}) as well as in metal additive manufacturing (\cite{sames2016metallurgy, gu2021material}). More specifically, the microstructural heterogeneities that form during these printing processes play a central role in the emergence of printing-orientation-dependent mechanical properties in additively manufactured materials and components (\cite{martin20173d, raney2018rotational, debroy2018additive, telles2025spatially}). However, a comprehensive mechanistic understanding of the microstructural mechanisms governing mechanical anisotropy in additively manufactured materials remains elusive, especially under large deformation. This is primarily attributed not only to experimental difficulties associated with direct measurements of local mechanical properties at the length scales of the printed microstructures but also due to the coupling between multiple microscopic deformation mechanisms.

Material jetting offers the distinctive capability of simultaneously depositing multiple materials with high precision, as widely demonstrated by both commercial (e.g., Stratasys and 3D Systems; Inkbit and Mimaki) and customized (e.g., \cite{sitthi2015multifab, roach2019m4, buchner2023vision}) multi-material printers. This is achieved by rapidly solidifying microscale liquid resins through photopolymerization upon UV exposure (\cite{ligon2016toughening, park20223d}). In particular, multi-material jetting processes enable the precise fabrication of “digital materials” by depositing hard and soft photocurable base resins at prescribed spatial locations. Varying the mixing ratio of these base resins allows digital materials to exhibit hybrid mechanical properties between thermoplastic and rubbery polymers. Over the past decade, digital materials whose mechanical behaviors show thermoplastic to rubbery features (e.g., \cite{yi2006large, deschanel2009rate, cho2013dissipation, cho2013constitutive, cho2017deformation, lee2023polyurethane}) have been widely employed to fabricate physical prototypes for a wide range of applications. Examples include mechanically (\cite{zhao2025modular}), thermally (\cite{wu2016multi, li2023algorithmic}), chemically (\cite{mao20163d}), and electromagnetically (\cite{akbari2019multimaterial}) responsive materials and devices. The printing of digital materials has also been widely used in the areas of soft robotics and actuation (\cite{bartlett20153d, wang2017biorobotic}). Recently, it has enabled the design and manufacture of heterogeneous or cellular architected materials that exhibit unusual mechanical functionalities (\cite{cho2016engineering, liu2018failure,fernandes2021mechanically, reyes2022tuning, lee2024extreme}).

In parallel, the deformation and failure of these digital materials have been characterized through mechanical testing (\cite{wu2016multi, akbari2019multimaterial, zorzetto2020properties, abu2025experimental}) and continuum-based constitutive modeling (\cite{slesarenko2018towards, xiang2019physically, lee2024extreme, moon2025extreme}). More recently, physics-augmented neural networks have been utilized to gain a better understanding of the deformation mechanisms in these materials, which often encompass significant microstructural uncertainties (\cite{yang2025physics, KLEIN2026105900}). It should also be noted that the base materials have been widely reported to exhibit nearly isotropic mechanical behavior from small to large strains (\cite{zhang2016transversely, zerhouni2019numerically, lee2024size}). However, digital materials for which two base resins are deposited together have been shown to exhibit significantly orientation-dependent elastic stiffness, plastic strength and fracture toughness (\cite{bass2016exploring, lumpe2019tensile, ituarte2019design}). Furthermore, optical microscopy observations have revealed highly anisotropic microstructures characterized by well-aligned and elongated hard inclusions embedded within the soft surrounding matrix, or $\mathit{vice}$ $\mathit{versa}$ (\cite{mueller2017mechanical, zorzetto2020properties}). These observations further highlight the need to gain a fundamental understanding of mechanical anisotropy in these 3D-printed digital materials, which would be essential to exploit their potential fully in a wide range of engineering applications.

In this work, by combining mechanical experiments, microscopy analysis and micromechanical modeling, we address the microstructural mechanisms that govern the printing-orientation-dependent mechanical behavior of digital materials fabricated using a high-resolution multi-material 3D printer (Connex3 Objet260, Stratasys Inc.). To this end, we construct representative volume elements (RVEs) that capture the key microstructural features observed in optical microscopy images of digital materials, where elliptical hard domains with high aspect ratios are well aligned within the surrounding soft matrix. Through micromechanical analysis, we demonstrate that high-aspect-ratio elliptical hard inclusions give rise to the initial elastic anisotropy. More importantly, our micromechanical modeling results reveal that the interplay between buckling instability and plastic deformation in elliptical hard domains connected along the printing direction governs the significant mechanical anisotropy in digital materials under large strains. The influence of the stiffness ratio between hard and soft domains is further explored to examine the orientation-dependent mechanical resilience and energy dissipation under loading and unloading conditions. Quantitative agreement is found between experimental and micromechanical modeling results, demonstrating that the approach presented in this work can advance our understanding of the printing-orientation-dependent mechanical behaviors observed in multi-material, 3D-printed digital materials.

The paper is organized as follows. The printing-orientation-dependent large-strain mechanical behavior of representative 3D-printed digital materials is presented in Section \ref{Mechanical anisotropy at small to large levels of strain}. We then investigate the microstructural features of the digital materials through optical microscopy analysis in Section \ref{Microscopy characterization} and Appendix \ref{Appendix microscopy anaylsis}. A facile procedure to construct representative volume elements that mimic the realistic hard and soft microstructures observed in digital materials is then presented in Section \ref{Micromechanical modeling framework}. Furthermore, we compare the micromechanical modeling results with the corresponding experimental data, by which we demonstrate that well-aligned elliptical hard inclusions play a crucial role in the mechanical anisotropy in these digital materials under small to large strain levels. We then discuss further implications of the micromechanical modeling results on the mechanical resilience and energy dissipation capabilities of these digital materials in Section \ref{Discussion and further implications}. We close the paper by summarizing the main findings in Section \ref{Conclusion}. Detailed information on the constitutive models used in the hard and soft components in the micromechanical analysis is presented in Appendix \ref{appendix constitutive modeling}. Additional micromechanical modeling results of the digital materials are provided in Appendix \ref{appendix micromechanical modeling analysis}.

\section{Mechanical anisotropy at small to large strains}
\label{Mechanical anisotropy at small to large levels of strain}
\begin{figure}[b!]
    \centering
    \includegraphics[width=1.0\textwidth]{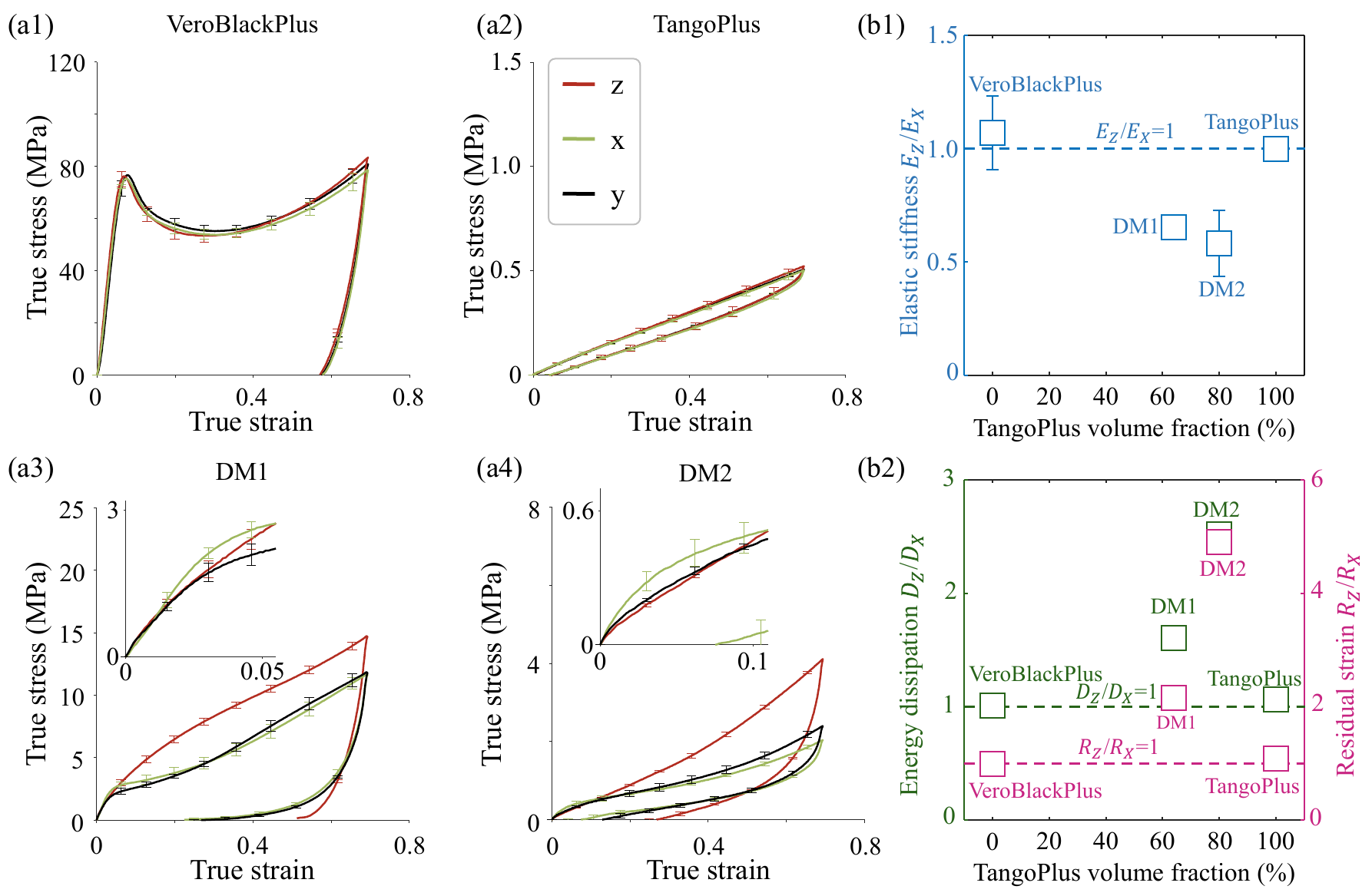}
    \caption{Printing-orientation-dependent mechanical behavior of 3D-printed materials. Stress-strain curves of (a1) and (a2) base materials (VeroBlackPlus and TangoPlus) and (a3) and (a4) digital materials (DM1 and DM2) printed in the x-, y- and z-directions during loading and unloading at a strain rate of 0.01 $\mathrm{s}^{-1}$ (insets: printing-orientation-dependent stress-strain curves of the DM1 and DM2 materials at small strains). (b1) Ratios of elastic stiffnesses and (b2) ratios of energy dissipations and residual strains in the digital materials printed in the z- and x-directions.}
    \label{fig:experiments}
\end{figure}
The orientation-dependent mechanical behavior of 3D-printed, digital materials is examined under large strain loading and unloading conditions. To this end, we selected two base materials, VeroBlackPlus$^{\mathrm{TM}}$ (hard) and TangoPlus$^{\mathrm{TM}}$ (soft), and their mixtures (i.e., digital materials) with Shore A hardness (\cite{StratasysDM}) of 95 and 70 (labeled DM1 and DM2 in this work). For each material, cylindrical specimens with a diameter of 10 mm and a height of 5 mm were printed in three different directions, each defined such that the axis of the global coordinate system (x, y, z) was aligned with the longitudinal axis of the cylindrical specimen. In this work, the global x- and z-axes were set to be the printing direction and the deposition direction of the sequential layers, respectively. After removing the support material (SUP705B$^{\mathrm{TM}}$) through manual brushing, the 3D-printed specimens were kept under ambient conditions for approximately 24 hours. We then conducted uniaxial compression tests on these specimens at a strain rate of 0.01 $\mathrm{s}^{-1}$ using a universal mechanical testing machine (INSTRON 4482) at room temperature (295K). Figure \ref{fig:experiments} presents the printing-orientation-dependent stress-strain curves of the base materials (VeroBlackPlus and TangoPlus) and the digital materials (DM1 and DM2). As shown in Figures \ref{fig:experiments}a1 and \ref{fig:experiments}a2, the base materials were found to be nearly isotropic at small to large strains. The hard VeroBlackPlus material exhibits standard thermoplastic polymeric behavior characterized by a relatively stiff initial elastic response followed by plastic yielding, strain softening and hardening due to chain alignments under large strain. Note that highly nonlinear unloading behavior is then observed, which results in a significant amount of energy dissipation as well as residual strain. The soft TangoPlus material displays hyperelastic rubbery behavior with negligible energy dissipation and hysteresis under cyclic loading conditions. In contrast, the digital materials, where the two base materials were mixed, were found to exhibit significant elastic and inelastic anisotropy during loading and unloading, as evidenced in Figures \ref{fig:experiments}a3 and \ref{fig:experiments}a4. Both the DM1 and DM2 materials printed along the x-direction show greater elastic stiffness than those printed along the y- and z-directions. Beyond the initial elastic regime, pronounced stress-rollover (or yield) and significantly lower flow stress levels are observed in both the DM1 and DM2 materials printed along the x- and y-directions. The unloading behavior of these digital materials is also shown to be strongly orientation-dependent. Specifically, upon unloading, the energy dissipation and residual strain were found to be greater when printing was conducted along the z-direction. The stress-strain curves of the digital materials printed along the z- and x-directions are further reduced to the ratios of the elastic stiffnesses, energy dissipation levels and residual strains as a function of the volume fraction of TangoPlus, as plotted in Figures \ref{fig:experiments}b1 and \ref{fig:experiments}b2. Given that the densities of VeroBlackPlus (1.17 $\sim$ 1.18 $\mathrm{g/cm^{3}}$) and TangoPlus (1.12 $\sim$ 1.13 $\mathrm{g/cm^{3}}$) are nearly identical (\cite{StratasysTangoPlus, StratasysVeroBlackPlus}), the estimated material consumption during the printing process of digital materials was directly used to determine the volume fractions of TangoPlus and VeroBlackPlus in these materials (e.g., \cite{slesarenko2018towards}). As shown in Figure \ref{fig:experiments}b1, both the DM2 ($\mathrm{E}_{\mathrm{z}}/\mathrm{E}_{\mathrm{x}}$ $\sim$ 0.6) and DM1 ($\mathrm{E}_{\mathrm{z}}/\mathrm{E}_{\mathrm{x}}$ $\sim$ 0.7) materials exhibit very strong (initial) elastic anisotropy, while the base materials (VeroBlackPlus and TangoPlus) are elastically isotropic ($\mathrm{E}_{\mathrm{z}}/\mathrm{E}_{\mathrm{x}}$ $\sim$ 1.0). The DM2 material also shows pronounced orientation dependence in both the energy dissipation and residual strain ($\mathrm{D}_{\mathrm{z}}/\mathrm{D}_{\mathrm{x}}$ $\sim$ 2.5 and $\mathrm{R}_{\mathrm{z}}/\mathrm{R}_{\mathrm{x}}$ $\sim$ 4.9), as presented in Figure \ref{fig:experiments}b2, while the DM1 material shows moderate inelastic anisotropy ($\mathrm{D}_{\mathrm{z}}/\mathrm{D}_{\mathrm{x}}$ $\sim$ 1.6 and $\mathrm{R}_{\mathrm{z}}/\mathrm{R}_{\mathrm{x}}$ $\sim$ 2.1). In contrast, the base materials were found to exhibit nearly isotropic mechanical resilience and energy dissipation ($\mathrm{D}_{\mathrm{z}}/\mathrm{D}_{\mathrm{x}}$ $\sim$ 1.0 and $\mathrm{R}_{\mathrm{z}}/\mathrm{R}_{\mathrm{x}}$ $\sim$ 1.0). These experimental results clearly show that mechanical anisotropy emerges in these multi-material, 3D-printed digital materials. More importantly, these digital materials exhibit markedly different mechanical anisotropy at small vs. large strains (see Figures \ref{fig:experiments}a3 and \ref{fig:experiments}a4); the digital materials printed along the deposition direction (z-direction) were found to exhibit lower initial elastic stiffness yet substantially higher flow stresses at increasing strains, accompanied by greater energy dissipation and residual strain upon unloading compared to those printed along the x- and y-directions; i.e., $\mathrm{E}_{\mathrm{x}} > \mathrm{E}_{\mathrm{z}}$ ($\mathrm{E}_{\mathrm{y}} \sim \mathrm{E}_{\mathrm{z}}$) at small strain while $\mathrm{D}_{\mathrm{x}} < \mathrm{D}_{\mathrm{z}}$ ($\mathrm{D}_{\mathrm{y}} < \mathrm{D}_{\mathrm{z}}$) and $\mathrm{R}_{\mathrm{x}} < \mathrm{R}_{\mathrm{z}}$ ($\mathrm{R}_{\mathrm{y}} < \mathrm{R}_{\mathrm{z}}$) under large strain, as clearly highlighted in Figures \ref{fig:experiments}b1 and \ref{fig:experiments}b2.

\section{Microscopy characterization of microstructures}
\label{Microscopy characterization}
\begin{figure}[b!]
    \centering
    \includegraphics[width=1.0\textwidth]{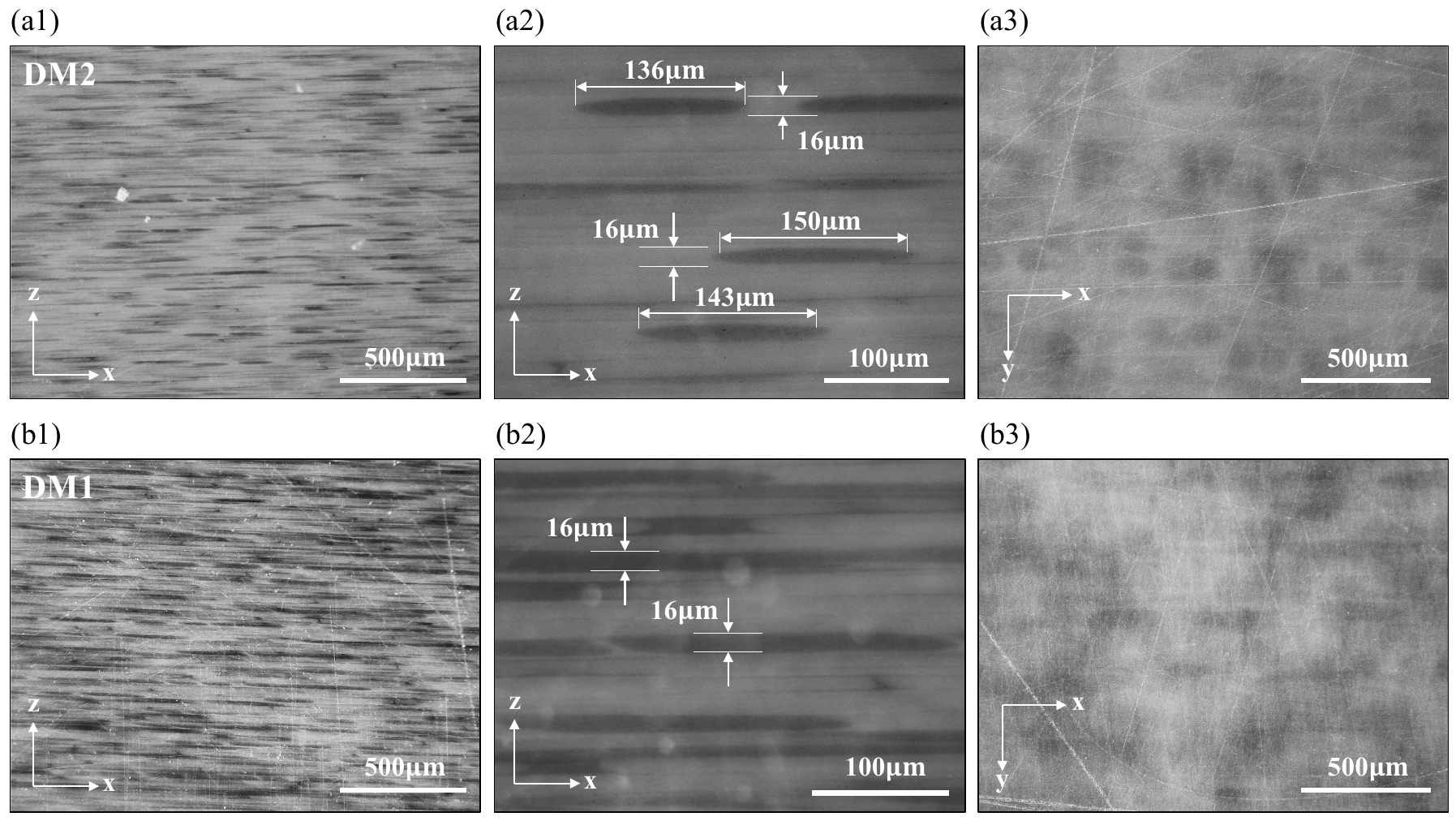}
    \caption{Microstructures of representative 3D-printed digital materials. Optical microscopy images of zx-plane cross-sections and the corresponding magnified views for (a1) and (a2) DM2 and (b1) and (b2) DM1 materials. (a3) and (b3) show the optical microscopy images of xy-plane cross-sections for the DM2 and DM1 materials, respectively.}
    \label{fig:optImage}
\end{figure}
We then examined the microstructural features underlying the digital materials (DM1 and DM2) using a high-resolution optical microscope (HRX-01, HiRox). To this end, each digital material specimen was mounted in a hard acrylic resin using a cylindrical mold and subsequently polished (MECATECH 250 SPI, Presi) with abrasive papers of progressively finer grit sizes; initial grinding was performed with 120-grit paper, followed by polishing with 400-, 800-, 1200-, 2400-, and 4000-grit papers to obtain an optically smooth surface suitable for an optical microscopy analysis. Figure \ref{fig:optImage} presents representative optical microscopy images of the cross-sections on the zx- and xy-planes of the DM1 and DM2 materials. The relatively hard and soft domains in these digital materials (mixtures of VeroBlackPlus and TangoPlus) appear as the dark and bright regions in the images, respectively. Due to the higher volume fraction of VeroBlackPlus, the dark regions in the DM1 materials appear broader than those in the DM2 materials. Importantly, as shown in Figures \ref{fig:optImage}a1 and \ref{fig:optImage}b1, the zx-plane cross-sectional images reveal well-aligned elliptical hard domains with high aspect ratios embedded within the surrounding soft matrix, in good agreement with optical microscopy images reported in previous studies (\cite{mueller2017mechanical, zorzetto2020properties}). The length of the major axis of the elliptical hard domains was found to be approximately 130 - 160 $\upmu$m, as depicted in Figure \ref{fig:optImage}a2, which is greater than the nominal resolution of the 3D printer in the x-direction ($\sim$ 40 $\upmu$m; \cite{StratasysObjet260Connex3}). These microstructural features are mainly attributed to the combined effects of diffusion between the two photocurable base resins (VeroBlackPlus and TangoPlus), as well as roller leveling in each layer during the multi-material jetting process (\cite{mueller2017mechanical, zorzetto2020properties}). Given that these base resins share constituent monomers, they readily mix and diffuse into one another (\cite{mueller2017mechanical, lei20183d}). Moreover, the length of the minor axis was found to be close to 16 $\upmu$m, nearly identical to the nominal resolution in the z-direction (\cite{StratasysObjet260Connex3}), as shown in Figures \ref{fig:optImage}a2 and \ref{fig:optImage}b2. This is mainly due to the layer-by-layer curing process as the printing progresses, during which the base resins are deposited onto a previously photocured layer, resulting in negligible diffusion and mixing across the printed layers (\cite{lei20183d, zorzetto2020properties, de2022modelling}). It should also be noted that due to the high volume fraction of VeroBlackPlus ($\sim$ 36\%), elliptical hard domains overlap significantly in the zx-plane of the DM1 materials, as shown in Figure \ref{fig:optImage}b2. Furthermore, hard domains with elliptical cross-sections of moderate aspect ratios are observed in the xy-plane images of the digital materials, as shown in Figures \ref{fig:optImage}a3 and \ref{fig:optImage}b3. However, clearly distinguishing the hard domains from the surrounding soft matrix was difficult due to the high optical transparency of the TangoPlus material as well as the small thickness of each printed layer. As an alternative, we fabricated specimens consisting of an array of VeroBlackPlus pillars (elongated along the z-direction) embedded in a TangoPlus matrix, as displayed in Figure \ref{fig:VBPillars} in Appendix \ref{Appendix microscopy anaylsis}. Each VeroBlackPlus pillar was set to have a square cross-section of 40 $\upmu$m X 40 $\upmu$m on the xy-plane, corresponding to the nominal resolution of the 3D printer in the x- and y-directions. However, as shown in Figure \ref{fig:VBPillars}d in Appendix \ref{Appendix microscopy anaylsis}, the elliptical cross-sections of the VeroBlackPlus pillars were found to have average major and minor axis lengths of approximately 152 $\upmu$m ($\pm$ 36 $\upmu$m) and 95 $\upmu$m ($\pm$ 25 $\upmu$m), respectively. This observation further supports that significant diffusion and mixing between the photocurable base resins occurs within each layer during the multi-material printing process. Microscopy characterization of the 3D-printed digital materials reveals the key microstructural features essential for constructing a representative volume element for micromechanical analysis. In particular, the orientation-dependent diffusion and mixing between the photocurable base resins leads to the formation of strongly anisotropic microstructures comprising two distinct domains.

\section{Micromechanical analysis}
\label{Micromechanical modeling framework}
\subsection{Construction of representative volume elements}
\begin{figure}[t!]
    \centering
    \includegraphics[width=1.0\textwidth]{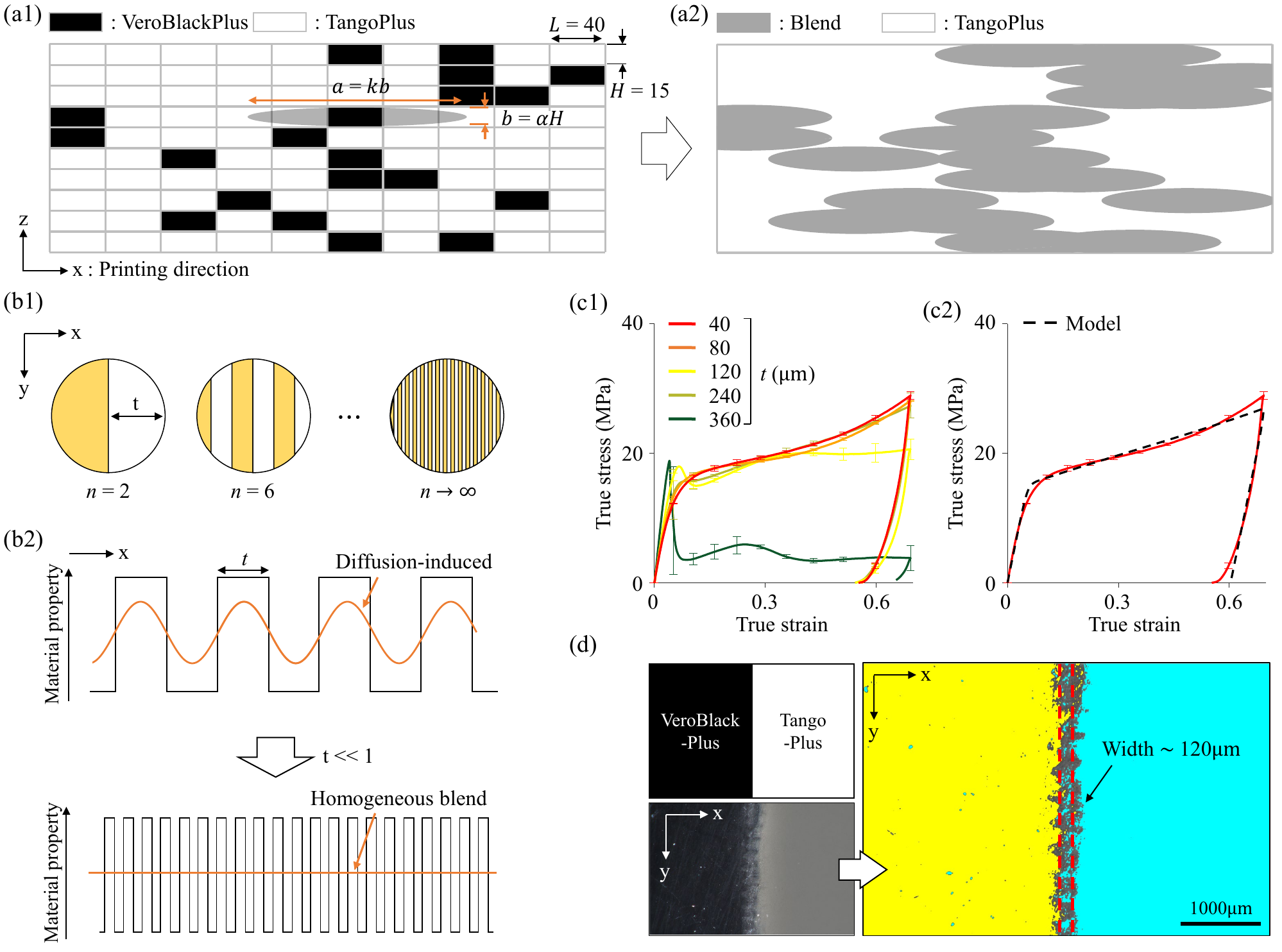}
    \caption{Construction of representative volume elements (RVEs) for 3D-printed digital materials. (a1) Schematic of the deposition pattern of the photocurable base resins (VeroBlackPlus and TangoPlus) on the zx-plane together with (a2) the corresponding RVE comprising well-aligned, high-aspect-ratio elliptical hard domains embedded within the surrounding soft matrix. (b1) Schematic illustration of cylindrical specimens comprising $n$ alternating VeroBlackPlus/TangoPlus layers along the x-direction with a layer thickness $t$; (b2) as the number of VeroBlackPlus/TangoPlus layers ($n$) increases (or the layer thickness $t$ decreases), a macroscopic response close to that of the homogeneous blend can be obtained due to diffusion and mixing between the photocurable base resins. (c1) Stress-strain curves at a strain rate of 0.01 $\mathrm{s}^{-1}$ for the cylindrical specimens with varying layer thicknesses ($t$ = 40, 80, 120, 240 and 360 $\upmu$m). (c2) Comparison of the macroscopic stress-strain responses measured experimentally (solid line) and predicted numerically (dashed line) for cylindrical specimens with $t$ = 40 $\upmu$m. (d) A broad, blurred interfacial region across VeroBlackPlus and TangoPlus layers along the x-direction.}
    \label{fig:rveDesign}
\end{figure}
Next, we present a simple procedure for designing a representative volume element (RVE) that captures the key microstructural features of the digital materials. We note that the RVEs are mainly taken to be two-dimensional for the micromechanical analysis; i.e., we construct RVEs that mimic the microstructures observed in the zx-cross-sections of the digital materials (e.g., see Figure \ref{fig:optImage}). The micromechanical modeling results for these two-dimensional RVEs are then presented together with experimental data for plane-strain compression of the digital materials in Figures \ref{fig:FLX2095-DM}, \ref{fig:FLX2070-DM} and \ref{fig:stiffnessRatio}. Furthermore, under uniaxial compressive loading conditions, we examine the printing-orientation-dependent mechanical behavior by constructing three-dimensional RVEs, especially for the DM2 materials in Figure \ref{fig:3DRVE}.

Figure \ref{fig:rveDesign}a1 schematically presents the deposition pattern of the two base resins on the zx-plane; note that the unit cells corresponding to the deposition locations of VeroBlackPlus and TangoPlus resins are displayed in black and white, respectively. Here, the length, $L$, and the height, $H$, of each cell are taken to be the nominal resolutions of the 3D-printer along the x- ($\sim$ 40$ \upmu$m) and z- ($\sim$ 15$ \upmu$m) directions. In each layer, VeroBlackPlus cells are randomly selected based on the volume fraction of VeroBlackPlus ($\mathrm{v}_{\mathrm{VB}}$), as estimated from the consumption of the VeroBlackPlus material during the printing process of the digital materials. For example, as shown in Figure \ref{fig:rveDesign}a1, two VeroBlackPlus cells out of ten cells are selected when constructing the RVEs for the DM2 materials with $\mathrm{v}_{\mathrm{VB}}$ $\sim$ 20\%. Then, to account for diffusion and mixing between the photocurable base resins, the selected VeroBlackPlus cells are replaced by ellipses placed at their centers, as schematically illustrated in Figures \ref{fig:rveDesign}a1 and \ref{fig:rveDesign}a2. The minor axis length of the ellipse, $b$, is defined as $b={\alpha}H$, where $\alpha$ is introduced to account for interlayer diffusion and mixing. The major axis length of the ellipse, $a$, is then defined as $a = kb$, where $k$ is the aspect ratio due to both diffusion and roller leveling within each layer. We further note that the elliptical “hard” domains are modeled as a homogeneous $50\%/50\%$ blend of VeroBlackPlus and TangoPlus while the remaining “soft” domains are taken to be pure TangoPlus; i.e., all VeroBlackPlus resins are assumed to diffuse and be mixed with the TangoPlus resins of equal mass in each case. Under this assumption, the total mass fraction of the elliptical hard domains is twice the estimated VeroBlackPlus consumption during the printing process of the digital materials. By also assuming that the densities of the two base materials and their blends are nearly identical, the corresponding “effective” volume fraction of the elliptical hard domains is taken to be $\mathrm{v}_{\mathrm{hard}} = 2\mathrm{v}_{\mathrm{VB}}$\footnote{Note that the densities of VeroBlackPlus and TangoPlus are 1.17 $\sim$ 1.18 $\mathrm{g/cm^{3}}$ and 1.12 $\sim$ 1.13 $\mathrm{g/cm^{3}}$, respectively.}. In other words, these elliptical hard domains are assumed not only to exhibit a softer response than the hard base material (VeroBlackPlus) but also to occupy a larger volume fraction effectively within each of the RVEs. Using $\mathrm{v}_{\mathrm{hard}}$, the pairs of $k$ and $\alpha$ for the elliptical hard domains are then simply determined. To characterize the "effective" stress-strain responses of this homogeneous blend (i.e., mixed VeroBlackPlus and TangoPlus domains in gray in Figure \ref{fig:rveDesign}a2), we fabricated cylindrical specimens comprising $n$ alternating VeroBlackPlus/TangoPlus layers along the x-direction, each with a thickness of $t$, as depicted in Figure \ref{fig:rveDesign}b1. Here, it was further assumed that when the number of layers ($n$) becomes sufficiently large, a macroscopic response close to that of the homogeneous blend or mixed domain can be obtained due to diffusion and mixing between the photocurable base resins, as schematically illustrated in Figure \ref{fig:rveDesign}b2. Mechanical tests of these specimens with varying layer thicknesses ($t$ = 40, 80, 120, 240 and 360 $\upmu$m) were then conducted under uniaxial compression along the deposition direction (z-direction) at a strain rate of 0.01 $\mathrm{s^{-1}}$. As displayed in Figure \ref{fig:rveDesign}c1, when $t$ $\geq$ 120 $\upmu$m, the specimens exhibit stress-strain responses that significantly vary with the layer thickness, indicating that homogeneous mixing of the base resins was not achieved in these specimens at this point. In particular, for the specimens with $t$ = 360 $\upmu$m, thin and discrete VeroBlackPlus layers are formed between the interfacial regions where diffusion and mixing occurs. Consequently, these specimens with $t$ = 360 $\upmu$m show much greater elastic stiffness, followed by a sudden stress drop due to the structural collapse of these thin VeroBlackPlus layers under compression. In contrast, the stress-strain responses of the specimens with $t$ = 40 $\upmu$m and 80 $\upmu$m are nearly identical at small to large strain levels. These results suggest that these layers ($t$ $<$ 80 $\upmu$m) are sufficiently thin to yield a homogeneous blend through diffusion and mixing between the base resins. The corresponding model prediction of the stress-strain behavior of this blend is shown along with the representative experimental data (for $t$ = 40 $\upmu$m) in Figure \ref{fig:rveDesign}c2; see Appendix \ref{appendix constitutive modeling} for detailed information of the finite deformation constitutive models for the homogeneous blend and pure TangoPlus. This approach for estimating the effective constitutive behavior of the hard domains is further supported by the optical microscopy images of a specimen comprising two alternating VeroBlackPlus/TangoPlus layers along the x-direction shown in the left insets of Figure \ref{fig:rveDesign}d. A broad, blurred interfacial region is observed on the xy-plane images of the specimen, as shown in the right inset of Figure \ref{fig:rveDesign}d. Because the alternating VeroBlackPlus and TangoPlus layers are stacked along the printing direction, the boundaries of the interfacial region appear wavy rather than straight, consistent with optical microscopy images reported in previous studies (\cite{liu2020effect}). Importantly, the width of the interfacial region ($\sim$ 120 $\upmu$m) is much greater than the layer thicknesses of the specimens, which exhibit nearly identical stress-strain responses ($t$ = 40 $\upmu$m and 80 $\upmu$m), further supporting that the large-strain mechanical behavior of these specimens can be used to model the elliptical hard domains in RVEs.

\subsection{Experiments vs. micromechanical models}
\label{Experiments vs. micromechanical models}
\begin{figure}[b!]
    \centering
    \includegraphics[width=1.0\textwidth]{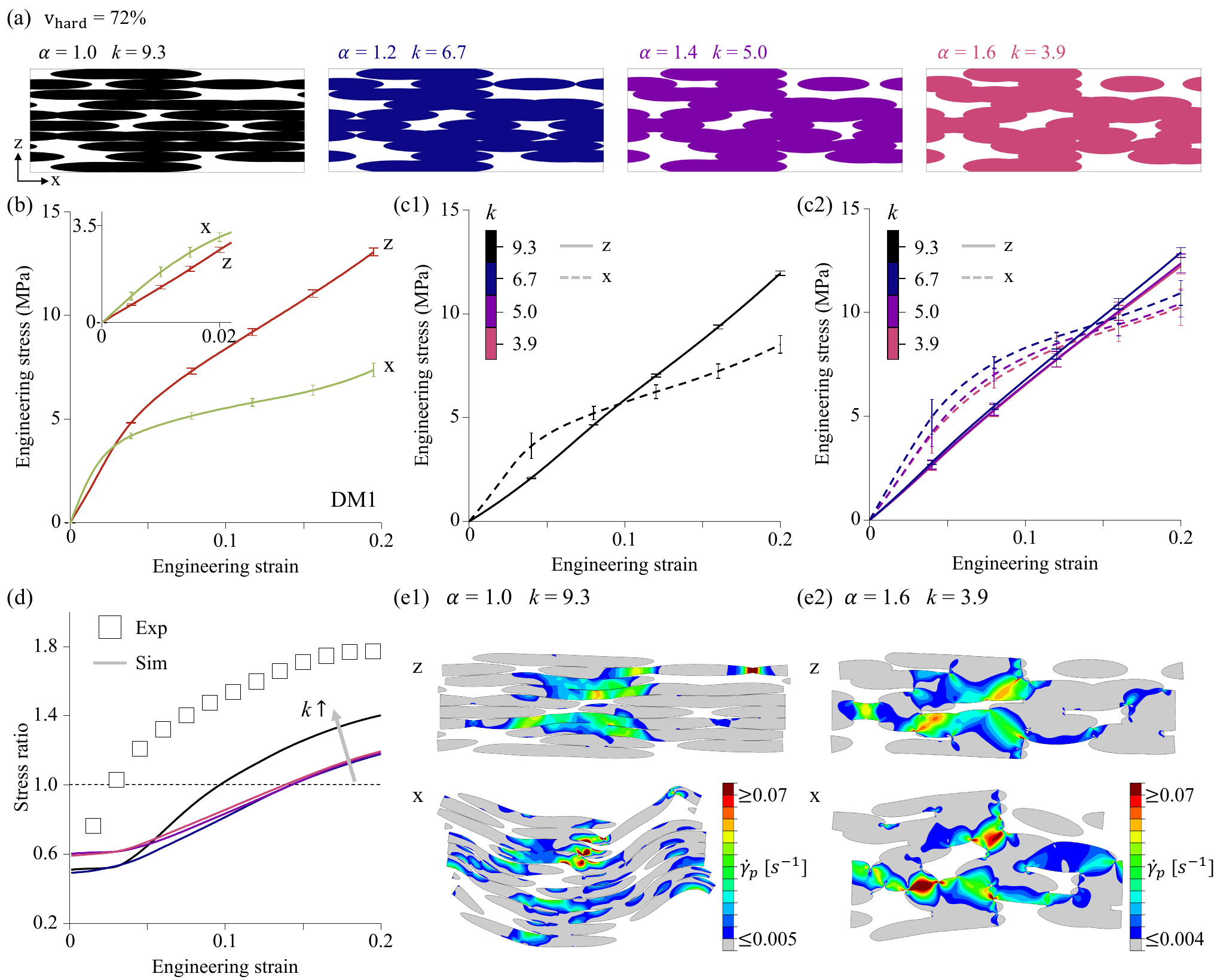}
    \caption{Experiments vs. micromechanical modeling results for the DM1 materials. (a) Representative volume elements for the DM1 materials ($\mathrm{v}_{\mathrm{hard}}$ $\sim$ 72\%) comprising elliptical hard domains with $k$ = 9.3, 6.7, 5.0 and 3.9. (b) Stress-strain curves (experiments) of the DM1 materials subjected to plane-strain compression along the global z- and x-axes at a strain rate of 0.01 $\mathrm{s}^{-1}$. Orientation-dependent mechanical behavior of RVEs (numerical simulations) with (c1) $k$ = 9.3 and (c2) $k$ = 6.7, 5.0 and 3.9. (d) Stress ratios, defined as the stress along the z-direction relative to that along the x-direction in experiments (open symbols) and numerical simulations (solid lines). Contours of the plastic shear strain rate $\dot{\gamma}_{p}$ in the deformed configurations of hard domains within RVEs with (e1) $k$ = 9.3 and (e2) $k$ = 3.9 loaded in the z- (upper insets) and x-directions (lower insets) at a macroscopic strain level of 0.15; here, only elliptical hard domains are displayed.}
    \label{fig:FLX2095-DM}
\end{figure}
Figure \ref{fig:FLX2095-DM} presents the micromechanical modeling results of the RVEs along with the experimental data for the DM1 materials. Here, $\alpha$ was set to 1.0, 1.2, 1.4 and 1.6, representing the degree of interlayer diffusion and mixing. For each $\alpha$, the corresponding aspect ratio $k$ due to the combined effects of diffusion between the TangoPlus and VeroBlackPlus resins and roller leveling in each layer was determined such that $\mathrm{v}_{\mathrm{hard}}$ across five statistical realizations of the RVEs was approximately 72\% (note that $\mathrm{v}_{\mathrm{VB}}$ $\sim$ 36\% for the DM1 materials). Consequently, as depicted in Figure \ref{fig:FLX2095-DM}a, the RVEs comprising the elliptical hard domains with $k$ = 9.3, 6.7, 5.0 and 3.9 were obtained. As $k$ increases (or with a decrease of $\alpha$), the connectivity of these hard domains increases in the x-direction but decreases in the z-direction. Also, due to the high $\mathrm{v}_{\mathrm{hard}}$ in these digital materials, the well-aligned elliptical domains are more likely to overlap along the x-direction, particularly in the RVEs with $k$ = 9.3 and $\alpha$ = 1.0, as observed in the microscopy images of the DM1 materials on the zx-plane (see Figure \ref{fig:optImage}). A micromechanical analysis of these RVEs was then conducted under macroscopic plane-strain compression along the z- and x-directions. Boundary value problems of the RVEs were solved using Abaqus/Standard. The macroscopic average responses of the RVEs were computed under periodic boundary conditions using the fictitious node virtual work method. Additionally, we used four-node bilinear elements with reduced integration and hourglass control (CPE4R). Furthermore, in order to examine the predictive capability of the proposed two-dimensional micromechanical models, plane-strain compression tests of the 3D-printed digital materials were additionally conducted at a strain rate of 0.01 $\mathrm{s}^{-1}$. To this end, specimens were placed in a fixture that restricts deformation in the out-of-plane direction (y-direction) and were subjected to compressive loading along the global z- or x-axes; we also followed the standard experimental protocols that have been widely used to characterize the large deformation behavior of polymeric materials under plane-strain conditions (\cite{arruda1993evolution}). As plotted in Figure \ref{fig:FLX2095-DM}b, the stress-strain responses of the DM1 materials were found to be highly orientation-dependent at small to large strains, very similar to those observed under uniaxial compression (see Figure \ref{fig:experiments}a3). It should be noted that the digital materials loaded in the x-direction exhibit higher initial elastic stiffness ($\mathrm{E}_{\mathrm{x}} > \mathrm{E}_{\mathrm{z}}$) yet substantially lower flow stress levels when under large strains than in the z-direction. Interestingly, the micromechanical modeling results of the RVEs with the highest aspect ratio ($k$ = 9.3 and $\alpha$ = 1.0) nicely capture the key features of the experimentally measured stress-strain curves of the DM1 materials at small to large strain levels, as shown in Figure \ref{fig:FLX2095-DM}c1. It is also important to note that the aspect ratios of the well-aligned elliptical hard domains observed in the optical microscopy images were found to be approximately 8 - 10 (see Figure \ref{fig:optImage}). The RVEs with smaller aspect ratios (i.e., $k$ = 6.7, 5.0 and 3.9) exhibit more gradual stress-rollover that occurs at considerably higher strain levels than with $k$ = 9.3 when loaded in the x-direction, as shown in Figure \ref{fig:FLX2095-DM}c2. These stress-strain curves with varying aspect ratios in the elliptical hard inclusions are reduced further to normalized stress, defined as the stress along the z-direction relative to that along the x-direction. As plotted in Figure \ref{fig:FLX2095-DM}d, under small strains ($<$ 0.03), the stress ratios remain below 1.0 due to the greater “initial” elastic stiffness when loaded in the x-direction. The stress ratios increase near the strain levels at which stress-rollover occurs in this loading direction. In particular, due to the pronounced rollover in the stress-strain curves of the RVEs with the highest aspect ratio, $k$ = 9.3, the stress ratio in these RVEs exhibits a dramatic upturn followed by a substantial increase under large strains, which is also clearly observed in the experimental data for the DM1 materials.

The anisotropic responses of the digital materials are also assessed by examining how the RVEs deform under the two loading directions. Figures \ref{fig:FLX2095-DM}e1 and \ref{fig:FLX2095-DM}e2 present the contours of the plastic shear strain rates $\dot{\gamma}_{p}$ of two exemplar RVEs with high and low aspect ratios ($k$ = 9.3 and 3.9) under macroscopic imposed strain of 0.15; the von Mises stress contours $\bar{\sigma}$ in these RVEs at a macroscopic strain level of 0.01 is presented in Figures \ref{fig:stressContour}a1 and \ref{fig:stressContour}a2 in Appendix \ref{appendix micromechanical modeling analysis}. At small strains, much greater stress develops throughout the elliptical hard domains when these domains are aligned with the loading direction (i.e., when loaded in the x-direction); the effects of well-aligned elliptical (or ellipsoidal) inclusions on elastic anisotropy have been widely reported in many classes of heterogeneous materials (\cite{sheng2004multiscale, saadat2015effective, raney2018rotational, mo2020microstructural}). However, as the macroscopic strain increases, the RVEs become mechanically unstable when loaded in the aligned direction of the high-aspect-ratio inclusions, as numerically predicted in the lower insets of Figures \ref{fig:FLX2095-DM}e1 and \ref{fig:FLX2095-DM}e2. More importantly, the RVEs with the highest aspect ratio ($k$ = 9.3) were found to undergo severe instability accompanied by local buckling of the elliptical hard domains (the lower inset of Figure \ref{fig:FLX2095-DM}e1). These micromechanical modeling results reveal that the apparent stress-rollover observed in the RVEs with $k$ = 9.3 loaded in the x-direction result from local buckling and severe plastic deformation of these hard domains. It should be noted that even with no significant local buckling, relatively gradual stress-rollover is observed in the RVEs with the lowest aspect ratio ($k$ = 3.9) loaded in the x-direction; see the stress-strain curve (the pink dashed line) in Figure \ref{fig:FLX2095-DM}c2 and the lower inset of Figure \ref{fig:FLX2095-DM}e2. This is primarily attributed not only to plastic deformation in the elliptical hard domains aligned well with the loading direction but to finite rotation of these domains in the soft matrices. Either local buckling or finite rotation of the elliptical hard domains in the RVEs loaded in the x-direction occurs because these deformation modes are energetically more favorable than the uniform compression of the hard domains under large deformation (\cite{moraleda2009finite, guttag2015locally, avazmohammadi2016macroscopic, zhao2016buckling, slesarenko2017microscopic}). In contrast, when loaded in the z-direction, both RVEs with $k$ = 9.3 and 3.9 deform with no instability, as displayed in the upper insets of Figures \ref{fig:FLX2095-DM}e1 and \ref{fig:FLX2095-DM}e2. Moreover, because the elliptical hard inclusions are perpendicular to the loading direction, they experience less significant plastic deformation, resulting in negligible stress-rollover and much higher stress responses at large strain levels, as shown in the simulated stress-strain curves (solid lines) in Figures \ref{fig:FLX2095-DM}c1 and \ref{fig:FLX2095-DM}c2. These results indicate that while the initial elastic anisotropy in the 3D-printed digital materials primarily results from stable compressive deformation of the high-aspect-ratio elliptical inclusions highly interconnected along the aligned direction, anisotropy under large strain is governed by local buckling as well as plastic deformation of these hard domains.

\begin{figure}[b!]
    \centering
    \includegraphics[width=1.0\textwidth]{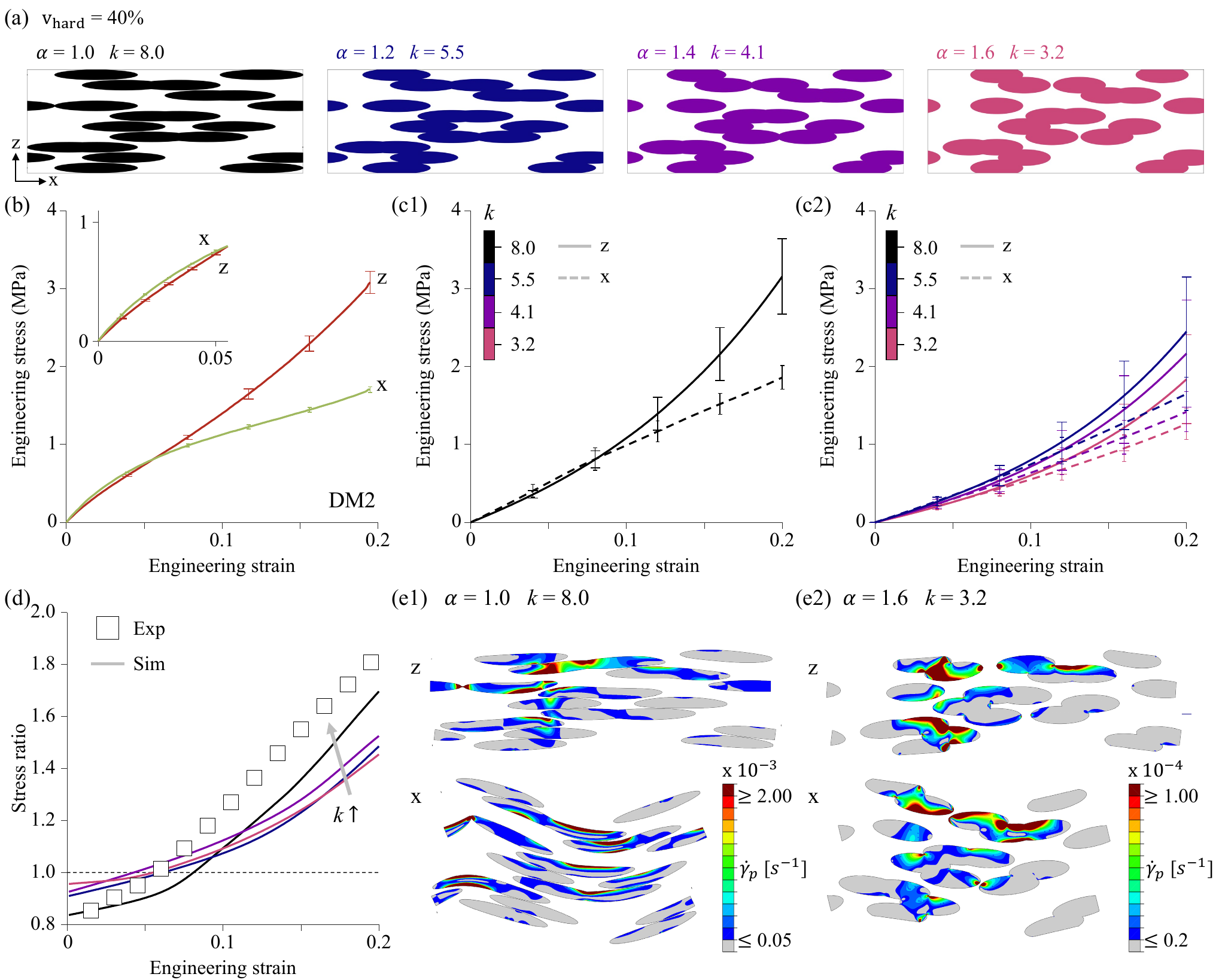}
    \caption{Experiments vs. micromechanical modeling results for the DM2 materials. (a) Representative volume elements for the DM2 materials ($\mathrm{v}_{\mathrm{hard}}$ $\sim$ 40\%) comprising elliptical hard domains with $k$ = 8.0, 5.5, 4.1 and 3.2. (b) Stress-strain curves (experiments) of the DM2 materials subjected to plane-strain compression along the global z- and x-axes at a strain rate of 0.01 $\mathrm{s}^{-1}$. Orientation-dependent mechanical behavior of RVEs (numerical simulations) with (c1) $k$ = 8.0 and (c2) $k$ = 5.5, 4.1 and 3.2. (d) Stress ratios, defined as the stress along the z-direction relative to that along the x-direction in experiments (open symbols) and numerical simulations (solid lines). Contours of the plastic shear strain rate $\dot{\gamma}_{p}$ in the deformed configurations of hard domains within RVEs with (e1) $k$ = 8.0 and (e2) $k$ = 3.2 loaded in the z- (upper insets) and x-directions (lower insets) at a macroscopic strain level of 0.15; here, only elliptical hard domains are displayed.}
    \label{fig:FLX2070-DM}
\end{figure}
The micromechanical modeling results are also presented in Figure \ref{fig:FLX2070-DM} of the RVEs of the DM2 materials. Here, $\mathrm{v}_{\mathrm{hard}}$ across five statistical realizations was set to approximately 40\%. RVEs with varying aspect ratios ($k$ = 8.0, 5.5, 4.1 and 3.2) were then constructed with the same set of $\alpha$ used for the RVEs of the DM1 materials (i.e., $\alpha$ = 1.0, 1.2, 1.4 and 1.6), as displayed in Figure \ref{fig:FLX2070-DM}a. Due to the relatively low $\mathrm{v}_{\mathrm{hard}}$, the elliptical hard domains are less connected (or overlapped) than in the RVEs for the DM1 materials. Figures \ref{fig:FLX2070-DM}b, \ref{fig:FLX2070-DM}c1 and \ref{fig:FLX2070-DM}c2 present the macroscopic stress-strain responses of the DM2 materials in experiments and numerical simulations. The RVEs with the highest aspect ratio ($k$ = 8.0 and $\alpha$ = 1.0) best match the experimental data. More specifically, when loaded in the x-direction, these RVEs exhibit greater initial elastic stiffness yet substantially lower flow stress levels under increasing strains. This is further supported by the stress ratio (the stress in the z-direction relative to that in the x-direction) which exceeds 1.0 with an increase in the macroscopic strain, after which it increases substantially under large strains, as plotted in Figure \ref{fig:FLX2070-DM}d. The mechanical anisotropy in these RVEs was found to decrease significantly with smaller aspect ratios ($k$ = 5.5, 4.1 and 3.2) from small to large strain levels. The numerically predicted stress-strain responses are examined further through the plastic shear strain rate and von Mises stress contours in the deformed configurations of two exemplar RVEs with $k$ = 8.0 and 3.2. At the early stage of loading, the alignment of the elliptical hard domains along the loading direction gives rise to greater elastic stiffness, especially in the RVEs with the highest aspect ratio ($k$ = 8.0), as clearly evidenced by the von Mises stress contours shown in Figures \ref{fig:stressContour}b1 and \ref{fig:stressContour}b2 in Appendix \ref{appendix micromechanical modeling analysis}. Beyond the initial elastic regime, buckling instabilities were found to emerge in the RVEs with $k$ = 8.0. However, the elliptical hard inclusions with $k$ = 3.2 exhibit finite rotation rather than local buckling, as shown in the lower insets of Figures \ref{fig:FLX2070-DM}e1 and \ref{fig:FLX2070-DM}e2. Furthermore, as shown in the upper insets of Figures \ref{fig:FLX2095-DM}e1 and \ref{fig:FLX2095-DM}e2, both RVEs with $k$ = 8.0 and $k$ = 3.2 deform without any instability when loaded in the z-direction. Interestingly, the stress ratio in the DM2 materials continues to increase significantly under large strains ($>$ $\sim$ 0.15), as evidenced in both the experiments and the numerical simulations (with $k$ = 8.0, solid black line), as presented in Figure \ref{fig:FLX2070-DM}d. However, the stress ratio in the DM1 materials (Figure \ref{fig:FLX2095-DM}d) was found not to increase yet to level off under large strain levels. This is primarily attributed to the large volume fraction of the hyperelastic soft matrices in the DM2 materials ($\mathrm{v}_{\mathrm{soft}}$ $\sim$ 60\%). When loaded especially in the z-direction, the DM2 materials undergo negligible plastic deformation but exhibit rubbery features characterized by a pronounced stress upturn due to chain orientation or locking at large strain levels; as shown in the upper insets of Figures \ref{fig:FLX2070-DM}e1 and \ref{fig:FLX2070-DM}e2, plastic flows with much smaller magnitudes develop throughout the elliptical hard inclusions compared to those in the DM1 materials with $\mathrm{v}_{\mathrm{hard}}$ $\sim$ 72\% (Figures \ref{fig:FLX2095-DM}e1 and \ref{fig:FLX2095-DM}e2). The comparisons of the experiments and numerical simulations presented in Figures \ref{fig:FLX2095-DM} and \ref{fig:FLX2070-DM} further demonstrate that the highly orientation-dependent mechanical behavior of these digital materials (DM1 and DM2) is nicely captured in the RVEs with $\alpha$ = 1.0. The micromechanical modeling results were also found to be very consistent with what we observed through the optical microscopy analysis, which revealed that interlayer diffusion and mixing is significantly suppressed, compared to intralayer diffusion and mixing between the photocurable base resins (VeroBlackPlus and TangoPlus) in each layer (e.g., see Figure \ref{fig:optImage}). More importantly, the intralayer diffusion and mixing resulted in higher connectivity between the hard elliptical inclusions along the printing direction (i.e., x-direction), by which these “connected” hard domains have much higher “effective” aspect ratios. Therefore, when loaded along the alignment direction, local buckling of these high-aspect-ratio hard inclusions becomes energetically more favorable than finite rotation, leading to the more pronounced stress-rollover observed in both the DM1 and DM2 materials. We note that buckling instabilities have also been experimentally observed in 3D-printed composites, where a soft base resin (TangoPlus) is used for the surrounding matrix and either a hard base resin (VeroBlackPlus) or digital materials (DM1 and DM2) is used for the high-aspect-ratio hard inclusions (\cite{li2013wrinkling, slesarenko2016harnessing, li2018instabilities, arora2022tunable, kaynia2025soft}).

\section{Discussion and further implications}
\label{Discussion and further implications}
Numerous micromechanical modeling strategies have been proposed to predict the effective elastic properties of additively manufactured materials with significant microstructural heterogeneities (\cite{mo2020microstructural, nawafleh2020additive, groetsch2023microscale}), including digital materials (\cite{meisel2018impact, de2022modelling}). Those micromechanical approaches have yet to fully capture the complex deformation mechanisms underlying the large strain mechanical behavior strongly dependent on the printing orientation. In this work, we have presented a methodology that combines large strain mechanical testing, microscopy characterization and micromechanical modeling to explore the mechanical anisotropy observed in 3D-printed digital materials at small to large strains. Representative volume elements have been proposed to mimic realistic anisotropic microstructures resulting from orientation-dependent diffusion and mixing between photocurable base resins during the printing process. The micromechanical modeling results of the RVEs presented in Figures \ref{fig:FLX2095-DM} and \ref{fig:FLX2070-DM} clearly show that the high-aspect-ratio elliptical hard domains play a central role in the emergence of mechanical anisotropy in the digital materials from small to large strains. Stable compressive deformation of the high-aspect-ratio elliptical inclusions interconnected along the printing direction was found to give rise to the “initial” elastic anisotropy, with $\mathrm{E}_{\mathrm{x}}/\mathrm{E}_{\mathrm{z}}$ $\sim$ 1.21 and 1.16 correspondingly for the representative DM1 and DM2 materials assessed experimentally along with $\mathrm{E}_{\mathrm{x}}/\mathrm{E}_{\mathrm{z}}$ $\sim$ 1.90 and 1.18 in the micromechanical modeling results of RVEs with $k$ = 9.3 and $k$ = 8.0, respectively. The interplay between local buckling and plastic deformation of the hard domains has been found to govern the mechanical anisotropy, especially under large strains.

\subsection*{$\bullet$ $\hspace*{0.5em}$ Resilience and dissipation of the digital materials upon unloading}
\begin{figure}[h!]
    \centering
    \includegraphics[width=1.0\textwidth]{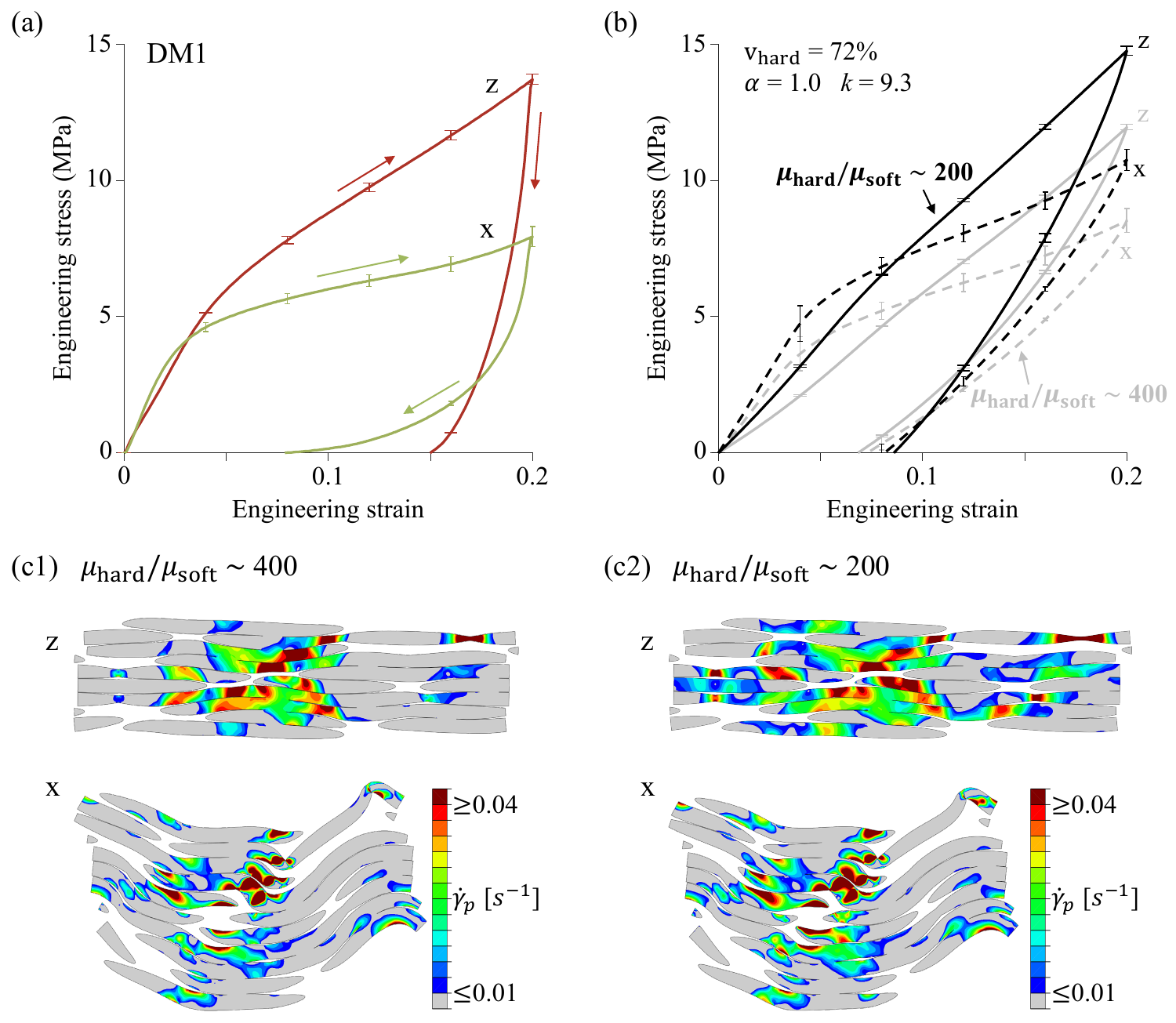}
    \caption{Printing-orientation-dependent mechanical resilience and energy dissipation of 3D-printed digital materials. (a) Stress-strain curves (experiments) of the DM1 materials loaded in the z- and x-directions under plane-strain conditions at a strain rate of 0.01 $\mathrm{s}^{-1}$. (b) Orientation-dependent mechanical behavior of RVEs (numerical simulations) for the DM1 materials ($k$ = 9.3 and $\alpha$ = 1.0) with $\mu_{\mathrm{hard}}$/$\mu_{\mathrm{soft}}$ $\sim$ 400 (gray lines) and $\mu_{\mathrm{hard}}$/$\mu_{\mathrm{soft}}$ $\sim$ 200 (black lines) loaded in the z- (solid lines) and x-directions (dashed lines). Contours of the plastic shear strain rate $\dot{\gamma}_{p}$ in the deformed configurations of hard domains in RVEs with (c1) $\mu_{\mathrm{hard}}$/$\mu_{\mathrm{soft}}$ $\sim$ 400 and (c2) $\mu_{\mathrm{hard}}$/$\mu_{\mathrm{soft}}$ $\sim$ 200 loaded in the z- (upper insets) and x-directions (lower insets) at a macroscopic strain level of 0.2; here, only elliptical hard domains are displayed.}
    \label{fig:stiffnessRatio}
\end{figure}
These digital materials also exhibit strongly orientation-dependent mechanical resilience and energy dissipation upon unloading, as demonstrated by plane-strain tests of the DM1 materials under loading and unloading conditions (Figure \ref{fig:stiffnessRatio}a). When loaded in the z-direction, greater energy dissipation ($\mathrm{D}_{\mathrm{z}}/\mathrm{D}_{\mathrm{x}}$ $\sim$ 1.3) and residual strain ($\mathrm{R}_{\mathrm{z}}/\mathrm{R}_{\mathrm{x}}$ $\sim$ 1.9) were observed. However, as shown in Figure \ref{fig:stiffnessRatio}b, the RVEs with $k$ = 9.3 and $\alpha$ = 1.0 loaded in the x-direction (dashed gray line) were found to exhibit greater energy dissipation and greater residual strain than those loaded in the z-direction (solid gray line; $\mathrm{D}_{\mathrm{z}}/\mathrm{D}_{\mathrm{x}}$ $\sim$ 0.87 and $\mathrm{R}_{\mathrm{z}}/\mathrm{R}_{\mathrm{x}}$ $\sim$ 0.94). The digital materials here were simply modeled as two-phase composites with a very large stiffness ratio between the hard and soft components ($\mu_{\mathrm{hard}}$/$\mu_{\mathrm{soft}}$), which resulted in a discrepancy between the experimental data and model predictions. It should also be noted that in the micromechanical analysis presented in Figures \ref{fig:FLX2095-DM} and \ref{fig:FLX2070-DM}, the stiffness ratio was approximately 400. In many previous studies of two-phase materials comprising hard inclusions embedded within surrounding soft matrices, such large stiffness ratios have been shown to degrade significantly the energy dissipation and load transfer capabilities through the hard inclusions (\cite{sheng2004multiscale, lee2024extreme, cho2024large}); i.e., most of the imposed macroscopic deformation is accommodated through surrounding matrices which are much softer than the inclusions. This is further supported by the contours of the plastic strain rates in the RVEs of the DM1 materials under macroscopic strain of 0.2. As shown in the upper inset of Figure \ref{fig:stiffnessRatio}c1, when loaded in the z-direction, significant plastic flows do not develop throughout the elliptical hard domains because the imposed macroscopic deformation is largely accommodated by the hyperelastic soft domains. Importantly, when loaded in the x-direction, the elliptical hard domains exhibit significant deformation through buckling, as shown in the lower inset of Figure \ref{fig:stiffnessRatio}c1. Consequently, strong plastic flows also develop throughout the buckled regions of these hard domains. Furthermore, when the RVEs deform with no instability at small strain, the elliptical inclusions aligned along the loading direction undergo more pronounced plastic deformation. Combined, these all contribute to the large amount of energy dissipation in the RVEs loaded in the x direction, resulting in $\mathrm{D}_{\mathrm{x}} > \mathrm{D}_{\mathrm{z}}$ in our micromechanical analysis. For further examination of the effect of the stiffness ratio on the printing-orientation-dependent energy dissipation capability, we conducted additional micromechanical analysis of the RVEs with $k$ = 9.3 and $\alpha$ = 1.0 by simply varying the elastic modulus of the soft component from $\mu_{\mathrm{soft}}$ = 0.26 MPa to 0.52 MPa (i.e., $\mu_{\mathrm{hard}}$/$\mu_{\mathrm{soft}}$ $\sim$ 200). As plotted in Figure \ref{fig:stiffnessRatio}b, both energy dissipation and residual strain were found to increase with the decreased stiffness ratio (i.e., from $\mu_{\mathrm{hard}}$/$\mu_{\mathrm{soft}}$ $\sim$ 400 to 200). Importantly, these RVEs with the lower stiffness ratio exhibited greater energy dissipation as well as greater residual strain when loaded in the z-direction; i.e., $\mathrm{D}_{\mathrm{x}}<\mathrm{D}_{\mathrm{z}}$ and $\mathrm{R}_{\mathrm{x}}<\mathrm{R}_{\mathrm{z}}$. This is further supported by the plastic strain rate contours in the elliptical hard domains. As shown in the upper inset of Figure \ref{fig:stiffnessRatio}c2, when loaded in the z-direction, significantly larger regions of the elliptical hard domains undergo plastic deformation, compared to those in the RVEs with $\mu_{\mathrm{hard}}$/$\mu_{\mathrm{soft}}$ $\sim$ 400 (Figure \ref{fig:stiffnessRatio}c1). In contrast, in the RVEs loaded in the x-direction, plastic flows develop primarily within the buckled hard domains under large strain, as shown in the lower inset of Figure \ref{fig:stiffnessRatio}c2, similar to the RVEs with $\mu_{\mathrm{hard}}$/$\mu_{\mathrm{soft}}$ $\sim$ 400 (Figure \ref{fig:stiffnessRatio}c1). It should also be noted that when these RVEs deform with no instability at small strain, decreasing the stiffness ratio results in much stronger plastic flows that develop throughout the hard inclusions perpendicular to the loading direction. However, because buckling instabilities occur at very low strain levels ($<$ 0.05), decreasing the stiffness ratio does not significantly influence the energy dissipation and load transfer capabilities in these RVEs loaded in the x-direction. These results also suggest that the interplay between buckling instability and plastic deformation of the well-aligned elliptical hard domains governs not only the mechanical anisotropy under large strain but also the orientation-dependent resilience and energy dissipation capabilities in these 3D-printed digital materials.

\subsection*{$\bullet$ $\hspace*{0.5em}$ Micromechanical modeling results of three-dimensional RVEs subjected to uniaxial compression}
\begin{figure}[h!]
    \centering
    \includegraphics[width=1.0\textwidth]{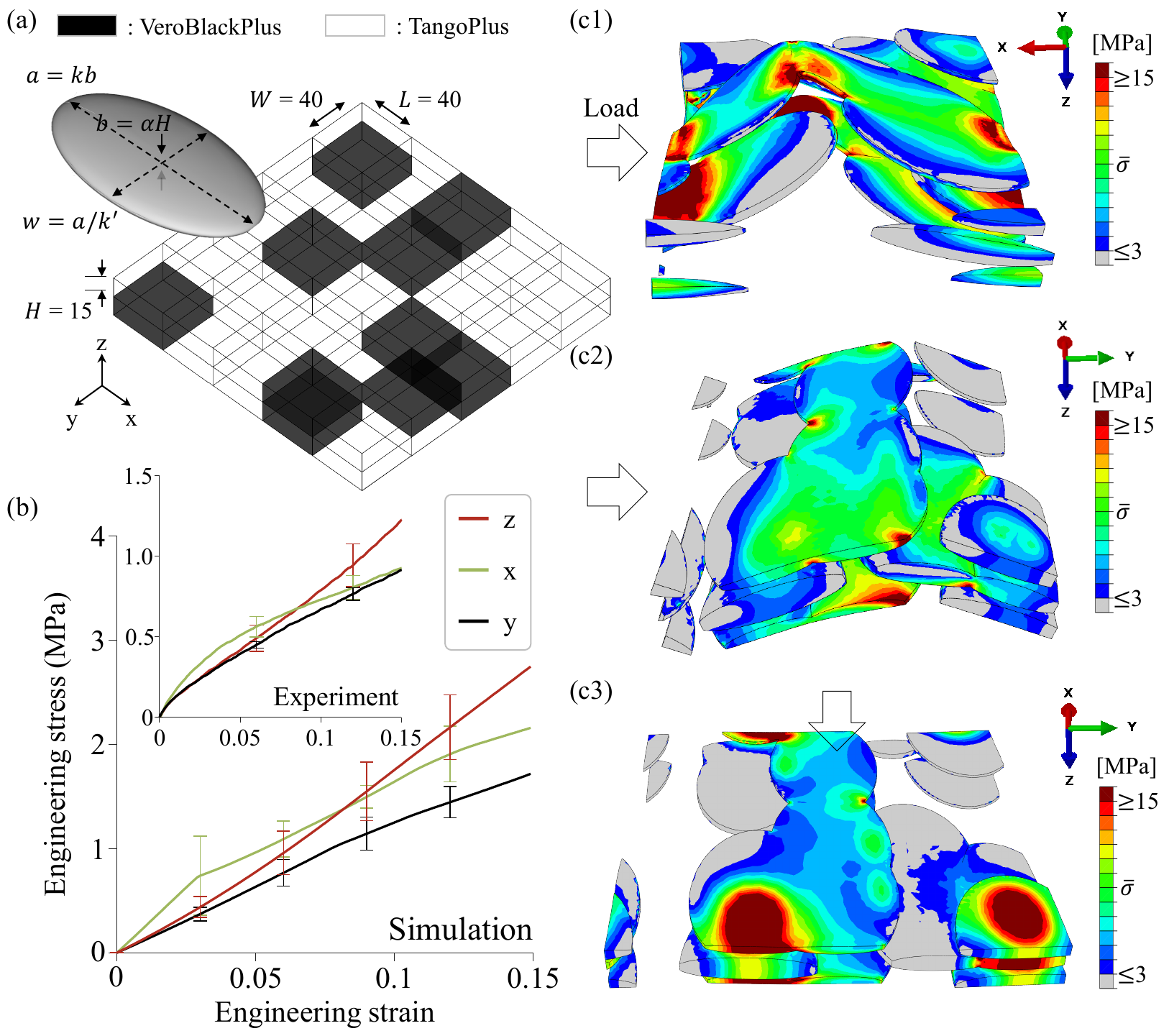}
    \caption{Micromechanical modeling results of three-dimensional representative volume elements (RVEs). (a) Schematic of the three-dimensional deposition pattern of the photocurable base resins (VeroBlackPlus and TangoPlus) together with an ellipsoidal inclusion used to construct hard domains in the three-dimensional setting. (b) Orientation-dependent stress-strain curves of three-dimensional RVEs for the DM2 materials (${\mathrm{v}}_{\mathrm{hard}}$ $\sim$ 40\%) with $k$ = 7.0, $k^{'}$ = 1.6 and $\alpha$ = 1.0 uniaxially compressed in the x-, y- and z-directions (inset: experimental stress-strain curves of the DM2 materials subjected to uniaxial compression along the global x-, y- and z-axes at a strain rate of 0.01 $\mathrm{s}^{-1}$). Contours of the von Mises stress ($\bar{\sigma}$) in the deformed configurations of hard domains in RVEs loaded in the (c1) x-, (c2) y- and (c3) z-directions at a macroscopic strain level of 0.15; here, only ellipsoidal hard domains are displayed.}
    \label{fig:3DRVE}
\end{figure}
We have presented micromechanical analysis of representative digital materials (DM1 and DM2) subjected to plane-strain compression using the two-dimensional RVEs in Figures \ref{fig:FLX2095-DM}, \ref{fig:FLX2070-DM} and \ref{fig:stiffnessRatio}. We note that relative deformations between the hard and soft domains through the out-of-plane direction (here, y-direction) do not play a crucial role in the overall mechanical features of the DM1 materials, as the volume fraction of the hard domains (${\mathrm{v}}_{\mathrm{hard}}$ $\sim$ 72\%) is relatively high (e.g., \cite{boyce2001micromechanisms, boyce2001micromechanics, cho2024large}). In contrast, for the DM2 materials with ${\mathrm{v}}_{\mathrm{hard}}$ $\sim$ 40\%, deformation through the out-of-plane direction may become important, which cannot be fully captured through a micromechanical analysis of two-dimensional RVEs. Here, the printing-orientation-dependent mechanical behavior of the DM2 materials is investigated further through a micromechanical analysis of three-dimensional RVEs subjected to uniaxial compression with no lateral geometric constraints; the micromechanical modeling results are compared with the uniaxial experimental data for the DM2 materials presented in Figure \ref{fig:experiments}a4.

As schematically illustrated in Figure \ref{fig:3DRVE}a, we started with three-dimensional deposition patterns of photocurable base resins (VeroBlackPlus and TangoPlus). Here, the width, $W$, of each cell is taken to be the nominal resolution of the 3D-printer along the y-direction ($\sim$ 40 $\upmu$m). From the VeroBlackPlus volume fraction in the DM2 materials (${\mathrm{v}}_{\mathrm{VB}}$ $\sim$ 20\%), five VeroBlackPlus cells were randomly selected in each layer of 5 by 5 cells on the xy-plane. These VeroBlackPlus cells were then replaced by ellipsoidal inclusions with dimensions $a$, $w$ and $b$ along the x-, y- and z-axes, respectively. The ellipsoidal dimension along the z-direction was set to $b={\alpha}H$, with $\alpha$ = 1.0 to account for the suppressed interlayer diffusion and mixing between the base resins (see Figure \ref{fig:optImage}). The aspect ratio of the elliptical cross-section of the ellipsoidal hard domains on the xy-plane was taken to be $k^{'}$ = 1.6 from what we observed in the specimens comprising VeroBlackPlus pillars embedded in a TangoPlus matrix, as shown in Figure \ref{fig:VBPillars} in Appendix \ref{Appendix microscopy anaylsis}. For $\alpha$ = 1.0 and $k^{'}$ = 1.6, the aspect ratio of the elliptical cross-section on the zx-plane, $k$, was determined such that the volume fraction of the ellipsoidal hard domains was ${\mathrm{v}}_{\mathrm{hard}}$ $\sim$ 40\%, similarly to the two-dimensional RVE construction presented in Figure \ref{fig:rveDesign}. We then conducted a micromechanical analysis of the RVEs constructed using three different deposition patterns under uniaxial compression along the x-, y- and z-directions; note that the material parameters used in the micromechanical analysis of two-dimensional RVEs were used here as well to model the hard inclusions and the surrounding soft matrices. Figure \ref{fig:3DRVE}b presents the macroscopic stress-strain responses of the three-dimensional RVEs for the DM2 materials. For comparison, experimentally measured stress-strain curves of the DM2 materials up to a strain of 0.15 are also displayed in the inset of Figure \ref{fig:3DRVE}b. In both the experiments and numerical simulations, the DM2 materials show macroscopic responses strongly dependent on the loading direction at small to large strain levels. Importantly, the numerical predictions of these three-dimensional RVEs reproduce the experimental results reasonably well. Under small strain, the RVEs loaded in the x-direction were found to exhibit greater initial elastic stiffness than those in the y- and z-directions ($\mathrm{E}_{\mathrm{x}}/\mathrm{E}_{\mathrm{z}}$ $\sim$ 1.8 and $\mathrm{E}_{\mathrm{y}}/\mathrm{E}_{\mathrm{z}}$ $\sim$ 0.9); i.e., $\mathrm{E}_{\mathrm{x}}>\mathrm{E}_{\mathrm{z}}$ and $\mathrm{E}_{\mathrm{y}} \sim \mathrm{E}_{\mathrm{z}}$. Because the well-aligned ellipsoidal hard domains are less connected along the y-direction, the elastic stiffening effect is weaker than that along the x-direction. Beyond the initial elastic regime, the stress-rollover is observed in the RVEs loaded not only in the x-direction but also in the y-direction. In particular, the RVEs loaded in the y-direction are found to exhibit less apparent stress-rollover than those loaded in the x-direction. This is further evidenced by the deformed configurations of these three-dimensional RVEs under macroscopic strain of 0.15. As shown in Figures \ref{fig:3DRVE}c1 and \ref{fig:3DRVE}c2, the RVEs loaded in the x- and y-directions become mechanically unstable with an increase in the macroscopic strain. More importantly, when loaded in the y-direction, the ellipsoidal hard inclusions predominantly undergo finite rotation (see Figure \ref{fig:3DRVE}c2) due to their lower effective aspect ratios on the yz-plane ($w/b$ = 4.375) compared to those on the zx-plane ($a/b$ = 7.0). However, as shown in Figure \ref{fig:3DRVE}c1 for the RVEs loaded in the x-direction, significant local buckling throughout the hard domains leads to more pronounced stress-rollover, similar to the two-dimensional RVEs with $k$ = 8.0 (see Figure \ref{fig:FLX2070-DM}e1). Furthermore, as shown in Figure \ref{fig:3DRVE}c3, when loaded in the z-direction, highly stable compressive deformation with negligible plastic deformation of the ellipsoidal hard domains results in negligible stress-rollover and much higher stress responses at large strains. The micromechanical modeling results of these three-dimensional RVEs further demonstrate that the two-dimensional approximation used for the DM2 materials with ${\mathrm{v}}_{\mathrm{hard}}$ $\sim$ 40\% presented in Figure \ref{fig:FLX2070-DM} is sufficient to capture their microscopic deformation mechanisms; i.e., when loaded in either the x- or z-direction, the overall deformation features are primarily governed by in-plane microstructural mechanisms rather than out-of-plane deformations. These three-dimensional RVEs with multiple ellipsoidal inclusions nicely capture the complex, anisotropic mechanical behavior of the DM2 materials from small to large strains under uniaxial conditions; however, employing RVEs with a larger number of ellipsoidal hard domains remains limited due to the high computational cost.

\section{Conclusion}
\label{Conclusion}
In this work, we have addressed, through a combination of mechanical experiments, microscopy analysis and micromechanical modeling, the deformation mechanisms responsible for the printing-orientation-dependent mechanical behavior of multi-material, 3D-printed digital materials, especially at large strains. Uniaxial and plane-strain compression tests clearly showed that mechanical anisotropy emerges in the 3D-printed digital materials under small to large strain levels. Specifically, these digital materials printed along the deposition direction (z-direction) have been found to exhibit lower initial elastic stiffness yet substantially higher flow stresses at increasing strains, accompanied by greater energy dissipation and residual strain upon unloading than those printed along the printing direction (x-direction). Microscopy characterization of the 3D-printed digital materials further revealed microstructural heterogeneities comprising two distinct hard and soft domains formed by orientation-dependent diffusion and mixing between photocurable base resins. Through a micromechanical analysis of representative volume elements that account for the key geometric features of the microstructures, we demonstrated that the proposed modeling framework can successfully predict the mechanical anisotropy of digital materials at small to large strains. At the early stage of deformation, stable compression of the high-aspect-ratio elliptical hard domains was shown to be responsible for the greater elastic stiffness in the digital materials when loaded in the printing direction. More importantly, our experimental and micromechanical modeling results showed that buckling instability coupled with plastic deformation of elliptical hard domains highly interconnected along the printing direction governs not only the mechanical anisotropy but the printing-orientation-dependent resilience and energy dissipation capabilities in these digital materials at large strains.

A comprehensive mechanistic understanding of the printing-orientation-dependent damage and fracture behavior of these 3D-printed digital materials remains beyond the scope of this work. More specifically, investigations on the toughening mechanisms in a broad range of 3D-printed digital materials with hard inclusions embedded in the surrounding soft matrices or $\mathit{vice}$ $\mathit{versa}$ would be of critical interest in future. Towards this end, a new computational methodology involving the phase-field or gradient-damage approach can be useful as these digital materials undergo large deformation prior to final failure (\cite{talamini2018progressive, narayan2021fracture, lee2023finite, lee2024size}). Further, the micromechanical modeling framework presented here can be extended to account for more realistic chemical compositions of these digital materials. The 3D-printed digital materials studied in this work have been found to have gradual variations in their composition and mechanical properties across the interfaces between hard and soft domains (\cite{mueller2017mechanical, liu2020effect, zorzetto2020properties}). These interfacial regions across hard and soft domains, also known as interphases, have been shown to influence not only the nonlinear elastic responses (\cite{qu2011nanoscale, goudarzi2015filled}) but also the load transfer and energy dissipation capabilities (\cite{ortiz2008bioinspired, li2020enhancing}). In our work, however, complex microstructural and chemo-mechanical features at the interphases were not considered; instead, sharp, perfectly bonded interfaces were assumed. While the present micromechanical models provide useful mechanistic insights into the mechanical anisotropy of these digital materials, particularly at large strains, they do not account for all major aspects of the experimentally measured responses. A more detailed characterization and modeling of interphases would enable more accurate predictions of the printing-orientation-dependent macroscopic responses of these digital materials. Finally, we note that most previous studies of architected materials (or mechanical metamaterials) were conducted, assuming that these 3D-printed digital materials are elastically and inelastically isotropic. Through the micromechanical modeling approaches presented in this work, which explicitly capture the printing-orientation-dependent mechanical features, the mechanical performance and functionalities of architected materials can be fully exploited from small to large strains.

\setcounter{figure}{0}
\setcounter{table}{0}
\setcounter{equation}{0}
\setcounter{section}{0}

\renewcommand{\theHfigure}{A.\arabic{figure}}
\renewcommand{\theHtable}{A.\arabic{table}}
\renewcommand{\theHequation}{A.\arabic{equation}}

\appendix
\renewcommand{\thesection}{Appendix~\Alph{section}}
\labelformat{section}{\Alph{section}}

\section{Further microscopy analysis on hard inclusions in a soft matrix in digital materials}
\label{Appendix microscopy anaylsis}
\renewcommand{\thefigure}{A.\arabic{figure}}
\renewcommand{\thetable}{A.\arabic{table}}
\renewcommand{\theequation}{A.\arabic{equation}}
\renewcommand{\thesubsection}{A.\arabic{subsection}}
\begin{figure}[h!]
    \centering
    \includegraphics[width=1.0\textwidth]{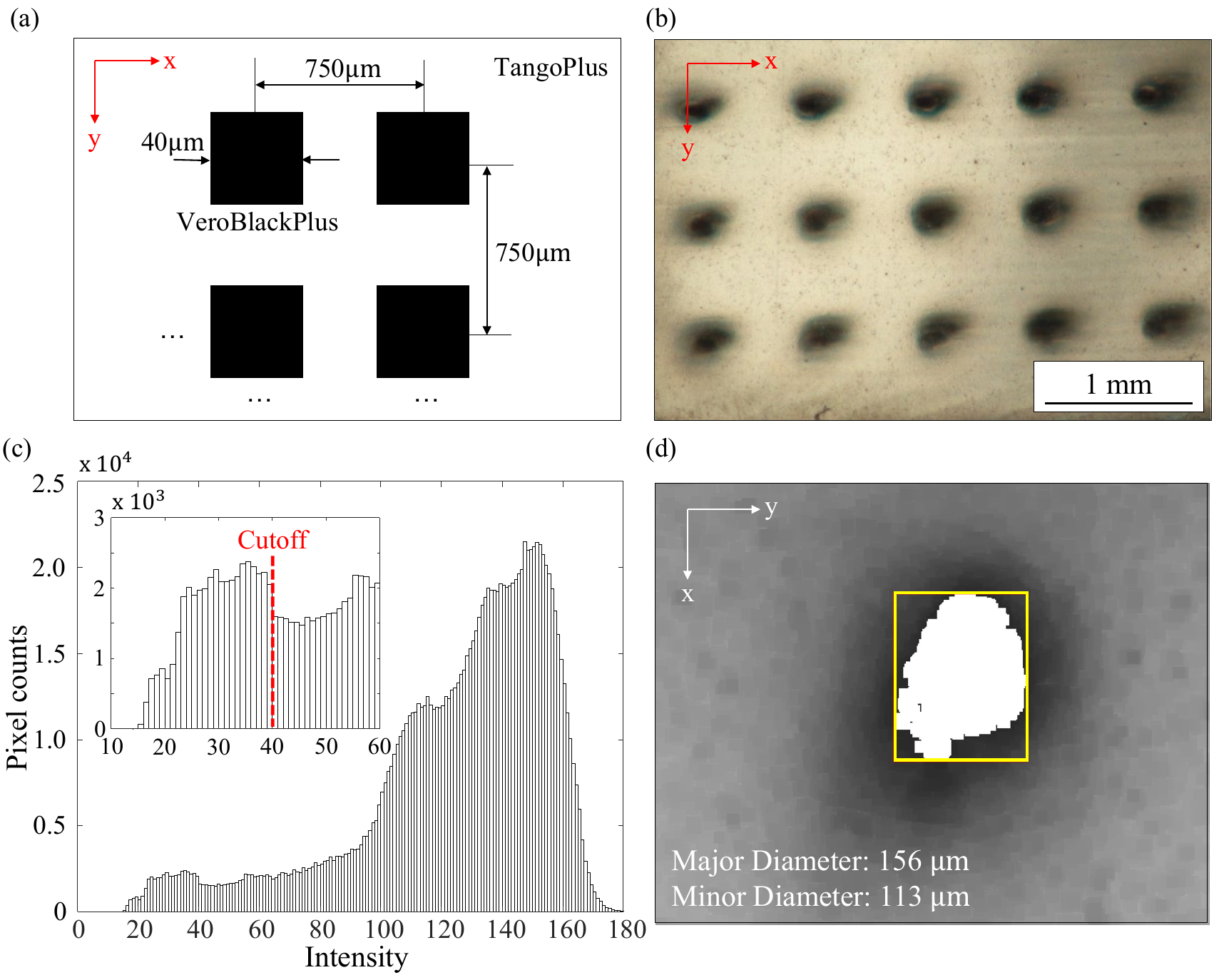}
    \caption{Orientation-dependent diffusion and mixing between photocurable base resins. (a) Schematic of an array of VeroBlackPlus pillars embedded within a TangoPlus matrix. (b) Optical microscopy image of an xy-plane cross-section of a 3D-printed specimen. (c) Intensity histogram of a representative grayscale image that shows a single hard domain embedded within the surrounding soft matrix (inset: magnified histogram of the low-intensity region together with the cutoff intensity used to distinguish dark and bright regions in the grayscale images). (d) Resulting binarized image of a hard domain with an elliptical cross-section on the xy-plane.}
    \label{fig:VBPillars}
\end{figure}
Diffusion and mixing between the two photocurable base resins (VeroBlackPlus and TangoPlus) during the printing processes presented in Figure \ref{fig:optImage} was further examined by fabricating specimens comprising an array of VeroBlackPlus pillars (elongated along the z-direction) embedded in a TangoPlus matrix, as schematically illustrated in Figure \ref{fig:VBPillars}a. A representative optical microscopy image of the xy-plane cross-section of these specimens is shown in Figure \ref{fig:VBPillars}b. For better preprocessing the raw optical images, only the red and green channels were used for grayscale conversion as the blue channel exhibited insufficient intensity contrast between the dark and bright regions. Specifically, the xy-plane cross-sections of the VeroBlackPlus pillars (dark regions in Figure \ref{fig:VBPillars}b) exhibited peak intensities of 38 - 40 across all RGB channels while the surrounding TangoPlus matrix (bright regions) showed peak intensities of 155 - 165 in the red and green channels but only about 125 in the blue channel. Furthermore, since the top and bottom surfaces of the specimens are not perfectly parallel, the VeroBlackPlus pillars appear slightly tilted with respect to the optical axis of the microscope (i.e., global z-axis). To avoid overestimating the xy-plane cross-sectional areas of the VeroBlackPlus pillars, a cutoff intensity of 40/255 was selected from the grayscale histogram, corresponding to the value at which the pixel count associated with the dark regions decreases sharply, as shown in Figure \ref{fig:VBPillars}c. As presented in Figure \ref{fig:VBPillars}d, binarization using this intensity threshold yields segmented pillar cross-sections that are well approximated by ellipses, with the average major and minor axis lengths of approximately 152 $\upmu$m ($\pm$ 36 $\upmu$m) and 95 $\upmu$m ($\pm$ 25 $\upmu$m), respectively. This additional microscopy analysis further supports our assumption that the elliptical (or ellipsoidal) hard domains exhibit an effectively larger volume fraction than ${\mathrm{v}}_{\mathrm{VB}}$ due to diffusion and mixing between photocurable base resins during the printing process.

\section{Mechanical behavior of hard and soft components: experiments and constitutive modeling}
\label{appendix constitutive modeling}
\renewcommand{\thefigure}{B.\arabic{figure}}
\renewcommand{\thetable}{B.\arabic{table}}
\renewcommand{\theequation}{B.\arabic{equation}}
\renewcommand{\thesubsection}{B.\arabic{subsection}}
\begin{figure}[t!]
    \centering
    \includegraphics[width=1.0\textwidth]{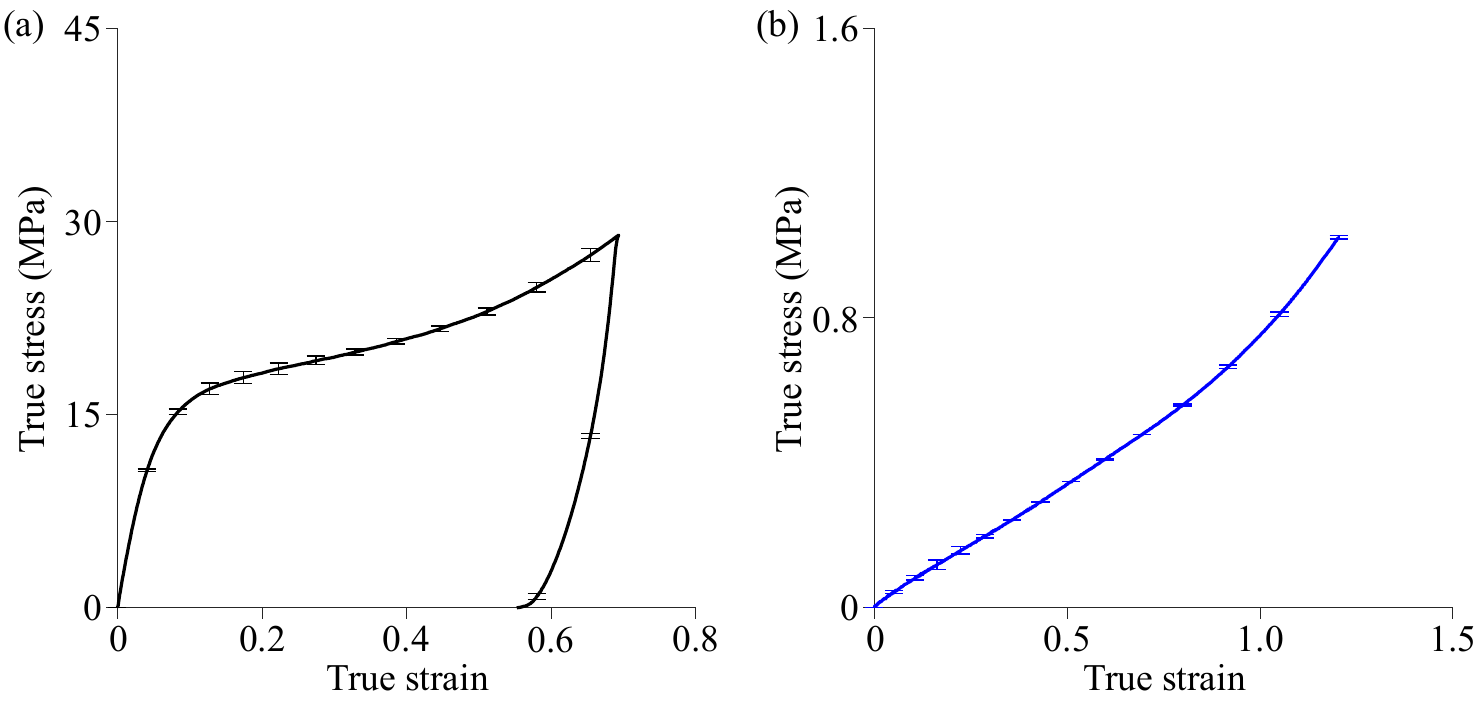}
    \caption{Stress-strain behavior of (a) the hard component and (b) the soft component under uniaxial compression at a strain rate of 0.01 $\mathrm{s}^{-1}$ in experiments.}
    \label{fig:hardSoftexp}
\end{figure}
Figure \ref{fig:hardSoftexp} presents the stress-strain behavior of the hard (a homogeneous 50\%/50\% blend of VeroBlackPlus and TangoPlus) and soft (pure TangoPlus) components under uniaxial compression at a strain rate of 0.01 $\mathrm{s}^{-1}$. The hard component exhibits a relatively stiff initial response, followed by yield-like stress rollover and strain hardening. The soft component shows a much more compliant stress-strain response.

We then present finite deformation constitutive modeling framework for the hard and soft components employed in the micromechanical analysis. The constitutive model of the hard component comprises (1) a time-dependent elastic-viscoplastic mechanism (denoted I) and (2) a time-independent hyperelastic mechanism (denoted N). We then define the following basic kinematic fields for the hard component:
\vspace*{0.3in}

\begin{tabular}{l l}
    $\mathbf{F} \defeq \frac{\partial \bm{\upvarphi}}{\partial \mathbf{X}} = \mathbf{F}_\mathrm{I} = \mathbf{F}_\mathrm{N}$ & deformation gradient that maps material points in a reference ($\mathbf{X}$) \\ & to points in a deformed configuration ($\mathbf{x} = \bm{\upvarphi}(\mathbf{X},t)$; $\bm{\upvarphi}$: motion) \\
    $\mathbf{F}_\mathrm{I} = \mathbf{F}^{e}_\mathrm{I} \mathbf{F}^{p}_\mathrm{I}$ & decomposition of $\mathbf{F}_\mathrm{I}$ into elastic ($\mathbf{F}^{e}_\mathrm{I}$) and plastic ($\mathbf{F}^{p}_\mathrm{I}$) parts; \\
    $\mathbf{F}^{e}_\mathrm{I} = \mathbf{R}^{e}_\mathrm{I} \mathbf{U}^{e}_\mathrm{I}$ & polar decomposition of $\mathbf{F}^{e}_\mathrm{I}$ into rotation ($\mathbf{R}^{e}_\mathrm{I}$) and stretch ($\mathbf{U}^{e}_\mathrm{I}$); \\
    $\bar{\mathbf{F}}_\mathrm{N} = J^{-1/3}\mathbf{F}_\mathrm{N}$ & isochoric part of $\mathbf{F}_\mathrm{N}$; \\
    $\bar{\mathbf{B}}_\mathrm{N} = \bar{\mathbf{F}}_\mathrm{N}\bar{\mathbf{F}}_\mathrm{N}^{\top}$ & isochoric left Cauchy-Green tensor. \\
\end{tabular}
\vspace*{0.3in}

\noindent The deformation rate is described by the velocity gradient $\mathbf{L} \defeq \mathrm{grad}\textbf{v}$ which is decomposed into elastic ($\mathbf{L}^{e}_{\mathrm{I}}$) and plastic ($\mathbf{L}^{p}_{\mathrm{I}}$) parts,
\begin{equation}
\begin{aligned}
\mathbf{L} & =\dot{\mathbf{F}}\mathbf{F}^{-1} \\
& =\dot{\mathbf{F}}^{e}_{\mathrm{I}}\mathbf{F}^{e-1}_{\mathrm{I}}+\mathbf{F}^{e}_{\mathrm{I}}\dot{\mathbf{F}}^{p}_{\mathrm{I}}\mathbf{F}^{p-1}_{\mathrm{I}}\mathbf{F}^{e-1}_{\mathrm{I}} \\
& = \mathbf{L}^{e}_{\mathrm{I}}+\mathbf{F}^{e}_{\mathrm{I}}\mathbf{L}^{p}_{\mathrm{I}}\mathbf{F}^{e-1}_{\mathrm{I}}.
\end{aligned}
\end{equation}

\noindent The plastic part of the velocity gradient is $\mathbf{L}^{p}_\mathrm{I} = \mathbf{D}^{p}_\mathrm{I} + \mathbf{W}^{p}_\mathrm{I}$, where $\mathbf{D}^{p}_\mathrm{I}$ is the rate of plastic stretching and $\mathbf{W}^{p}_\mathrm{I}$ is the plastic spin. Furthermore, we make two important kinematical assumptions for plastic flow; the flow is incompressible (i.e., $\operatorname{det}\mathbf{F}^p_{\mathrm{I}}=1$) and irrotational (i.e., $\mathbf{W}^{p}_\mathrm{I}=0$). Thus, the rate of change in the plastic deformation gradient is given by,
\begin{equation}
\begin{aligned}
        \dot{\mathbf{F}}^p_\mathrm{I} = \mathbf{D}^p_\mathrm{I} \mathbf{F}^{p}_\mathrm{I}.
\end{aligned}
\end{equation}

\noindent The Cauchy stress ($\mathbf{T}_{\mathrm{I}}$) in the time-dependent elastic-viscoplastic mechanism I is expressed by,
\begin{equation}
\begin{aligned}
\mathbf{T}_{\mathrm{I}} = \frac{1}{J} \mathbf{R}^{e}_{\mathrm{I}} \mathbf{M}^{e}_{\mathrm{I}} \mathbf{R}^{e \top}_{\mathrm{I}},
\end{aligned}
\end{equation}

\noindent with the Mandel stress $\mathbf{M}^{e}_{\mathrm{I}} = 2\mu_{\mathrm{I}} (\ln \mathbf{U}^{e}_{\mathrm{I}})_{0} \, + \, K(\ln J)\mathbf{I}$, the elastic shear modulus $\mu_\mathrm{I}$ and the bulk modulus $K$. We note that the total bulk response is lumped into the time-dependent mechanism I. The rate of plastic stretching $\mathbf{D}^p_\mathrm{I}$ is assumed to be coaxial to the deviatoric part of the Mandel stress (i.e., $(\mathbf{M}^e_\mathrm{I})_0 = \mathbf{M}^e_\mathrm{I} - \frac{1}{3}\mathrm{tr}(\mathbf{M}^e_\mathrm{I})\mathbf{I}$),

\begin{equation}
\mathbf{D}^{p}_{\mathrm{I}} =\frac{\dot{\gamma}^{p}_{\mathrm{I}}}{\sqrt{2}}\mathbf{N}^{p}_{\mathrm{I}}, \quad \text{where} \quad \mathbf{N}^{p}_{\mathrm{I}} = \frac{(\mathbf{M}^{e}_{\mathrm{I}})_{0}}{\|(\mathbf{M}^{e}_{\mathrm{I}})_{0}\|} \quad \text{and} \quad\|(\mathbf{M}^{e}_{\mathrm{I}})_{0}\| = \sqrt{(\mathbf{M}^{e}_{\mathrm{I}})_{0} : (\mathbf{M}^{e}_{\mathrm{I}})_{0}}.
\end{equation}
Then, we employ the thermally-activated viscoplasticity model prescribed by,
\begin{equation}
    \dot{\gamma}^{p}_{\mathrm{I}} = \dot{\gamma}_{\mathrm{I},0} \operatorname{exp}\left[ -\frac{\Delta G}{k_{\mathrm{B}}\theta} \left\{1-\frac{\bar{\tau}}{s_{\mathrm{0}}}  \right\}   \right], \quad \text{where} \quad \bar{\tau} = \frac{1}{\sqrt{2}} \|(\mathbf{M}^{e}_{\mathrm{I}})_{0}\|,
\end{equation}
with the reference plastic strain rate $\dot{\gamma}_{\mathrm{I},0}$, the activation energy $\Delta G$, Boltzmann's constant $k_{\mathrm{B}}$, the absolute temperature $\theta=295\mathrm{K}$ (room temperature), the shear strength $s_{\mathrm{0}}$ and the magnitude of the deviatoric part of the Mandel stress $\bar{\tau}$ (i.e., von Mises equivalent shear stress).

The (deviatoric) Cauchy stress ($\mathbf{T}_{\mathrm{N}}$) in the time-independent hyperelastic mechanism N is given by,
\begin{equation}
\mathbf{T}_{\mathrm{N}} = \frac{\mu_{\mathrm{N}}}{3J} \frac{\lambda_{\mathrm{N}}}{\bar{\lambda}} \mathscr{L}^{-1} \left(\frac{\bar{\lambda}}{\lambda_{\mathrm{N}}}\right) (\bar{\mathbf{B}}_{\mathrm{N}})_0 \quad \text{where} \quad \bar{\lambda} = \sqrt{\frac{\text{tr}\bar{\mathbf{B}}_{\mathrm{N}}}{3}} \quad \text{and} \quad (\bar{\mathbf{B}}_{\mathrm{N}})_0 = \bar{\mathbf{B}}_{\mathrm{N}} - \frac{1}{3}\mathrm{tr}(\bar{\mathbf{B}}_{\mathrm{N}})\mathbf{I},
\end{equation}
with the elastic shear modulus $\mu_{\mathrm{N}}$ and the limiting chain extensibility $\lambda_{\mathrm{N}}$; $\mathscr{L}^{-1}$ is the inverse Langevin function with $\mathscr{L}(x)=\coth(x)-\dfrac{1}{x}$.

\begin{table}[t!]
\centering
\small{
\begin{tabular}{lcc}
\hline
\textbf{Hard component: Time-dependent elastic-viscoplastic mechanism I} &  &  \\
\hline
& $\mu_{\mathrm{I}}$ [MPa] & 90.6 \\
$\mathbf{T}_{\mathrm{I}} = \frac{1}{J} \mathbf{R}^{e}_{\mathrm{I}} \mathbf{M}^{e}_{\mathrm{I}} \mathbf{R}^{e \top}_{\mathrm{I}}, \quad \text{where} \quad \mathbf{M}^{e}_{\mathrm{I}} = 2\mu_{\mathrm{I}} (\ln \mathbf{U}^{e}_{\mathrm{I}})_{0} \, + \, K(\ln J)\mathbf{I}$
& $K$ [GPa] & 1.17 \\
& $\Delta G$ [$10^{-20}$J] & 3.5 \\
$\dot{\gamma}^{p}_{\mathrm{I}} = \dot{\gamma}_{\mathrm{I},0} \operatorname{exp}\left[ -\frac{\Delta G}{k_{\mathrm{B}}\theta} \left\{1-\frac{\bar{\tau}}{s_{\mathrm{0}}}  \right\}   \right], \quad \text{where} \quad \bar{\tau} = \frac{1}{\sqrt{2}} \|(\mathbf{M}^{e}_{\mathrm{I}})_{0}\|$ & $\dot{\gamma}_{\mathrm{I},0}$ [$s^{-1}$] & 0.04 \\
& $s_{0}$ [MPa] & 9.04 \\
\hline
\textbf{Hard component: Time-independent hyperelastic mechanism N} &  &   \\
\hline
$\mathbf{T}_{\mathrm{N}} = \frac{\mu_{\mathrm{N}}}{3J} \frac{\lambda_{\mathrm{N}}}{\bar{\lambda}} \mathscr{L}^{-1} \left(\frac{\bar{\lambda}}{\lambda_{\mathrm{N}}}\right) (\bar{\mathbf{B}}_{\mathrm{N}})_0$ & $\mu_{\mathrm{N}}$ [MPa] & 5.4  \\
& $\lambda_{\mathrm{N}}$ & $\sqrt{6}$ \\
\hline
\textbf{Soft component} &  &   \\
\hline
& $\mu_{\mathrm{S}}$ [MPa] & 0.235  \\
$ \mathbf{T}_{\mathrm{Soft}}=\frac{\mu_\mathrm{S}}{3 J} \frac{\lambda_\mathrm{S}}{\bar{\lambda}} \mathscr{L}^{-1}\left(\frac{\bar{\lambda}}{\lambda_\mathrm{S}}\right)\left(\bar{\mathbf{B}}\right)_0 + {K_\mathrm{S}} (J-1) \mathbf{I}$ & $K_\mathrm{S}$ [MPa] & 16.71  \\
& $\lambda_\mathrm{S}$ & $\sqrt{6.5}$ \\
\hline
\end{tabular}
}
\caption{Material parameters used in the constitutive models for hard and soft components.}
\label{Tab:material_parameter}
\end{table}

\begin{figure}[b!]
    \centering
    \includegraphics[width=1.0\textwidth]{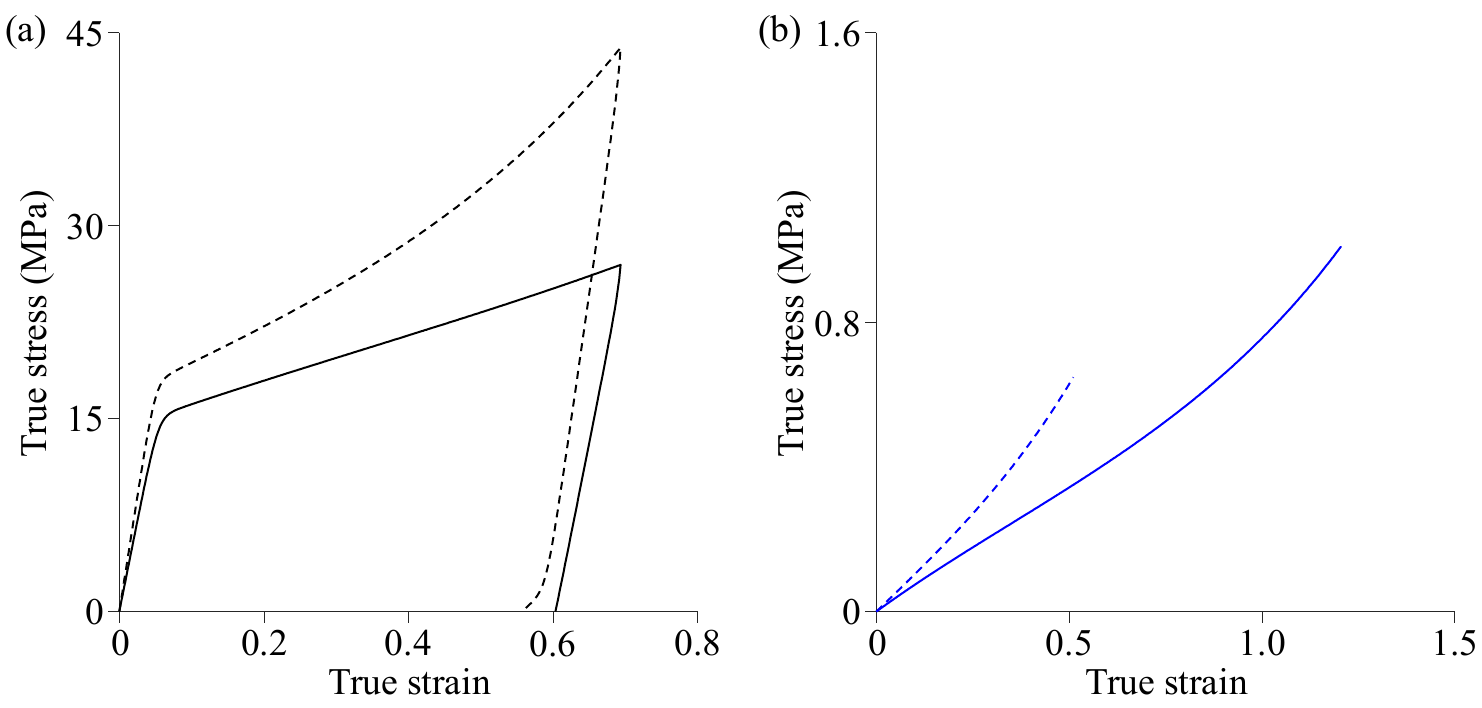}
    \caption{Stress-strain behavior of (a) the hard component (black lines) and (b) the soft component (blue lines) under uniaxial (solid lines) and plane-strain (dashed lines) compression at a strain rate of 0.01 $\mathrm{s}^{-1}$ in numerical simulations.}
    \label{fig:hardSoftsim}
\end{figure}

The total stress in the hard component is then obtained by,
\begin{equation}
\mathbf{T}_{\mathrm{hard}} = \mathbf{T}_{\mathrm{I}} + \mathbf{T}_{\mathrm{N}}.
\end{equation}

A nearly incompressible Arruda-Boyce model is used for the soft component. The Cauchy stress in the soft component is expressed by,
\begin{equation}
\begin{aligned}
    \mathbf{T}_{\mathrm{Soft}}=\frac{\mu_\mathrm{S}}{3 J} \frac{\lambda_\mathrm{S}}{\bar{\lambda}} \mathscr{L}^{-1}\left(\frac{\bar{\lambda}}{\lambda_\mathrm{S}}\right)\left(\bar{\mathbf{B}}\right)_0 + {K_\mathrm{S}} (J-1) \mathbf{I} \quad &\text{where} \quad \bar{\lambda} = \sqrt{\frac{\mathrm{tr}(\bar{\mathbf{B}})}{3}},
\end{aligned}
\end{equation}
where $\mu_\mathrm{S}$ is the elastic shear modulus, $K_\mathrm{S}$ is the bulk modulus, $\lambda_\mathrm{S}$ is the limiting chain extensibility, $\bar{\mathbf{B}} = J^{-2/3} \mathbf{F} \mathbf{F}^{\top}$, $\mathbf{F}$ is the deformation gradient and $\bar{\mathbf{B}}_0$ is the deviatoric part of the isochoric left Cauchy-Green tensor (i.e., $\bar{\mathbf{B}}_0 = \bar{\mathbf{B}} - \frac{1}{3}\mathrm{tr}(\bar{\mathbf{B}})\mathbf{I}$) in the soft component.

Figure \ref{fig:hardSoftsim} presents the stress-strain curves numerically simulated using the constitutive models for both hard and soft components under uniaxial (solid lines) and plane-strain (dashed lines) compression conditions. The material parameters used in the constitutive models are given in Table \ref{Tab:material_parameter}.

\section{Further micromechanical modeling results}
\label{appendix micromechanical modeling analysis}
\renewcommand{\thefigure}{C.\arabic{figure}}
\renewcommand{\thetable}{C.\arabic{table}}
\renewcommand{\theequation}{C.\arabic{equation}}
\renewcommand{\thesubsection}{C.\arabic{subsection}}
\begin{figure}[h!]
    \centering
    \includegraphics[width=1.0\textwidth]{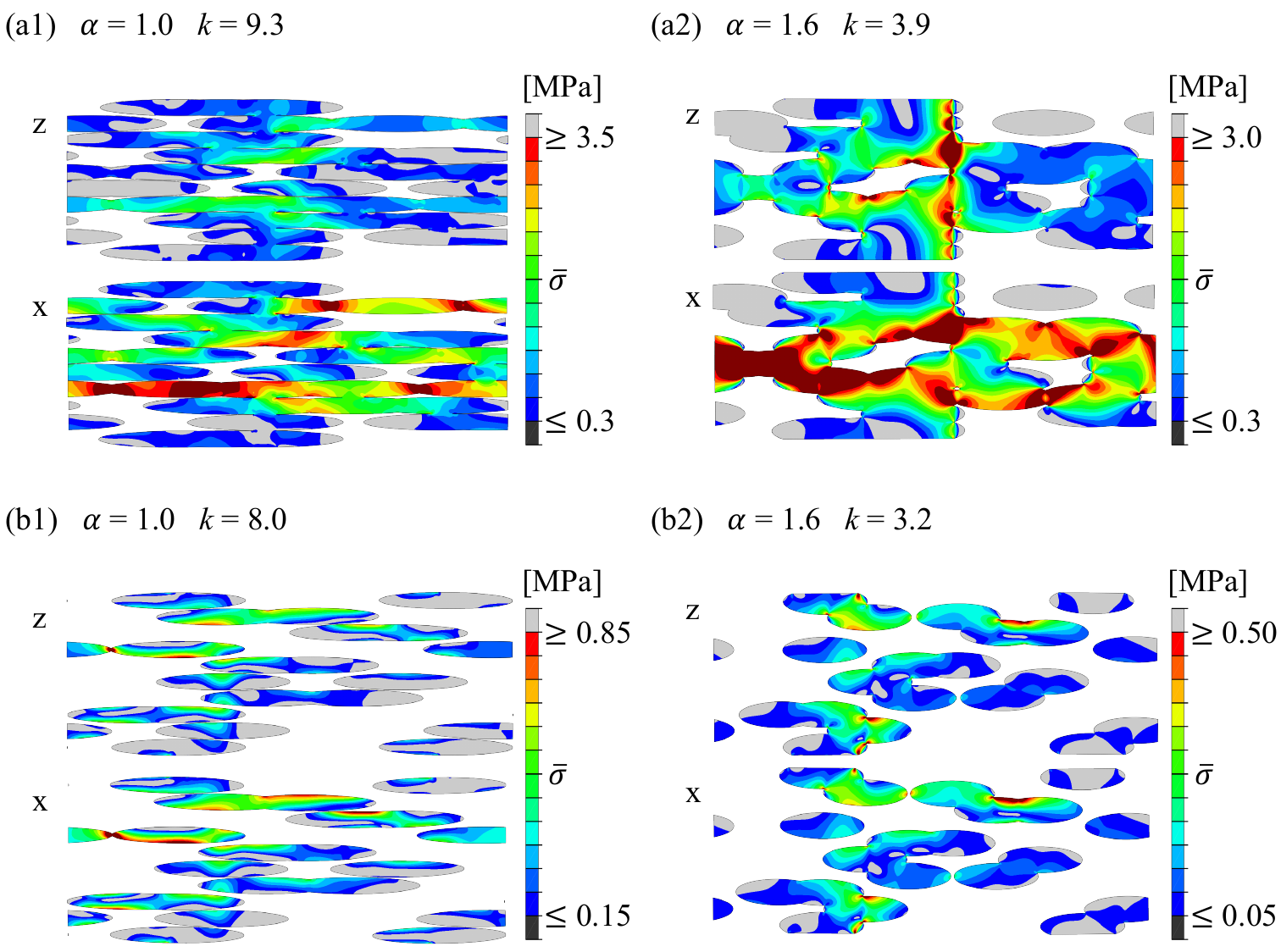}
    \caption{Contours of the von Mises stress ($\bar{\sigma}$) in the deformed configurations of hard domains in RVEs for the (a1) and (a2) DM1 and (b1) and (b2) DM2 materials loaded in the z- (upper insets) and x-directions (lower insets) at a macroscopic strain level of 0.01; here, only ellipsoidal hard domains are displayed.}
    \label{fig:stressContour}
\end{figure}
We here present the von Mises stress contours in the elliptical hard domains in the RVEs for the DM1 materials with $k$ = 9.3 (Figure \ref{fig:stressContour}a1) and 3.9 (Figure \ref{fig:stressContour}a2) at a macroscopic strain level of 0.01; and the RVEs for the DM2 materials with $k$ = 8.0 (Figure \ref{fig:stressContour}b1) and 3.2 (Figure \ref{fig:stressContour}b2). At these early stages of loading, all of the RVEs were found to deform without any instability. Importantly, much greater stress develops throughout the elliptical hard domains in the RVEs loaded along the alignment direction of these domains (x-direction) than those loaded in the z-direction. The von Mises stress contours further demonstrate that the influence of the well-aligned elliptical domains on “initial” elastic anisotropy becomes more pronounced in the RVEs with the highest aspect ratios; i.e., $k$ = 9.3 and 8.0 for the DM1 and DM2 materials, respectively.

\section*{Acknowledgement}
This work was supported by the National Research Foundation of Korea (RS-2023-00279843) and Korea Advanced Institute of Science and Technology (N11250018, N11250083).

\clearpage
\printbibliography
\end{document}